\documentclass[12pt]{article}
\usepackage{multirow,verbatim}
\usepackage{latexsym, amsbsy, amssymb}
\usepackage{natbib}
\usepackage{tikz}
\usepackage{graphicx}
\usepackage{epsfig}
\usepackage{amsmath}
\usepackage{mathrsfs}
\usepackage{array}
\usepackage{enumerate} % must be put before enumitem
\usepackage[shortlabels]{enumitem}
\usepackage{multirow}
\usepackage{latexsym, amsbsy, amssymb}
\usepackage{verbatim}
\usepackage{color}
\usepackage{url}
\usepackage{caption}
\usepackage{subcaption}
\usepackage{bbm}
\def\boe{\begin{enumerate}}
\def\eoe{\end{enumerate}}

\newtheorem{proposition}{{\bf Proposition}}
\newtheorem{lemma}{{\bf Lemma}}
\newtheorem{corollary}{{\bf Corollary}}
\newtheorem{assumption}{{\bf Assumption}}
\newtheorem{definition}{{\bf Definition}}
\newtheorem{theorem}{{\bf Theorem}}

\newtheorem{example}{{\bf Example}}

\newcommand\ca[1]{{\cal{#1}}}
\newcommand\lo[1]{_{\nano{#1}}}
\newcommand\hi[1]{^{\nano{#1}}}

\def\diag{{\mathrm{diag}}}

\def\argmin{\mathrm{argmin}}

\def\L{{\cal L}}

\def\trans{^{\mbox{\tiny{\sf  T}}}}

\def\inv{^{\mbox{\tiny $-1$}}}

\def\var{\mathrm{var}}
\def\cov{\mathrm{cov}}

\def\tr{\mathrm{tr}}

\def\eop{\hfill $\Box$ \\
}
\def\trans{^{\mbox{\tiny{\sf T}}}}

\def\ali{&\,}

\def\ran{\mathrm{ran}}

\def\of{{\nano {\circ}}}
\def\nano{\scriptscriptstyle}

\def\real{{\mathbb R}}

\def\ka{\kappa}

\def\L2T{L \lo 2 (T)}
\def\L2TX{L \lo 2 (T\lo X)}
\def\L2TX{L \lo 2 (T\lo Y)}

\def\ali{&\,}
\def\spn{{\rm{span}}}
\def\ali{&\,}

\def\eod{\end{document}}

\def\shortbar{\rule[-.0em]{.04em}{.15em}}
\def\bbar{\mbox{\raisebox{-.05em}{$\,\overset{\overset{\shortbar}{\shortbar}}{\shortbar}\,$}}}

\def\cran{{\overline{{\rm{ran}}}}}

\newcommand{\red}[1]{{\leavevmode\color{red}{#1}}}

\newfont{\rsfsten}{rsfs10 scaled 1100}
\newfont{\rsfstena}{rsfs10 scaled 800}
\newfont{\rsfstenb}{rsfs10 scaled 500}

\def\shortbar{\rule[-.0em]{.04em}{.15em}}
\def\bbar{\mbox{\raisebox{-.05em}{$\,\overset{\overset{\shortbar}{\shortbar}}{\shortbar}\,$}}}

\def\var{{\mathrm{var}}}

\def\eop{\hfill $\Box$ \\ }

\def\nano{\scriptscriptstyle}
\def\real{\mathbb R}
\def\var{\mathrm{var}}
\def\hii#1{^{\nano{(\mathrm{\uppercase{#1}})}}}

\def\ran{\mathrm{ran}}
\def\ker{\mathrm{ker}}
\def\cran{\overline{\mathrm{ran}}}

\def\cov{\mathrm{cov}}
\def\nano{\scriptscriptstyle}
\def\inv{\hi{\nano -1}}

\def\nano{\scriptscriptstyle}
\def\of{\mbox{\raisebox{1pt}{$\nano{\circ}$}}}
\def\ka{\kappa}

\def\ali{&\,}

\def\diag{\mathrm{diag}}

\def\hii#1{\hi{(#1)}}

\def\spn{\mathrm{span}}

\def\diag{\mbox{diag}}

\def\frechet{Fr\'echet     }

\def\onex{\mathbbm{1} \lo X}
\newcommand{\RN}[1]{%
	\textup{\uppercase\expandafter{\romannumeral#1}}%
}
\newcommand{\ben}{\begin{enumerate}}
	\newcommand{\een}{\end{enumerate}}

\begin{document}
	
\begin{center}
	{\Large{\bf
			Nonlinear global Fr\'echet regression for random objects via weak conditional expectation}\\
		\vskip.5cm
		Satarupa Bhattacharjee, Bing Li, and Lingzhou Xue}\\
	\vskip.3cm
	Department of Statistics, The Pennsylvania State University\\
	University Park, PA 16802, U.S.A.
	
\end{center}

	\begin{abstract}
		Random objects are complex non-Euclidean data taking value in general metric space, possibly devoid of any underlying vector space structure. Such data are getting increasingly abundant with the rapid advancement in technology. Examples include probability distributions, positive semi-definite matrices, and data on Riemannian manifolds. However, except for regression for object-valued response with Euclidean predictors and distribution-on-distribution regression, there has been limited development of a general framework for object-valued response with object-valued predictors in the literature. To fill this gap, we introduce the notion of a weak conditional Fr\'echet mean based on Carleman operators and then propose a global nonlinear Fr\'echet regression model through the reproducing kernel Hilbert space (RKHS) embedding. Furthermore, we establish the relationships between the conditional Fr\'echet mean and the weak conditional Fr\'echet mean for both Euclidean and object-valued data. We also show that the state-of-the-art global Fr\'echet regression recently developed  by \cite{pete:19} emerges as a special case of our method by choosing a linear kernel. We require that the metric space for the predictor admits a reproducing kernel, while the intrinsic geometry of the metric space for the response is utilized to study the asymptotic properties of the proposed estimates. Numerical studies, including extensive simulations and a real application, are conducted to investigate the performance of our estimator in a finite sample.
		\end{abstract}
%%%%%%%%%%%%%%%%%%%%%%%%%%%%%%%%%%%%%%%%%%%%%%
%% Please use \tableofcontents for articles %%
%% with 50 pages and more                   %%
%%%%%%%%%%%%%%%%%%%%%%%%%%%%%%%%%%%%%%%%%%%%%%
%\tableofcontents
\section{Introduction}
\label{intro}
Encountering complex non-Euclidean data, taking values in a general metric space that may defy any inherent linear structure, has become increasingly common in areas such as biological or social sciences with the rapid advancement of technology.
Examples of such ``\emph{random object}'' data, recorded in the form of images, shapes, networks, or life tables~\citep{marr:alon:14}
include distributional data in Wasserstein space~\citep{deli:17, le:17},
symmetric positive definite matrix objects~\citep{dryd:09},
data on the surface of the sphere~\citep{di:14},
phylogenetic trees~\citep{bill:01},	and  finite-dimensional Riemannian manifolds objects~\citep{afsa:11, bhat:03,bhat:05, penn:18, afsa:11,huck:15},
among others. Since the data are metric space valued, many classical notions of statistics, such as the definition of sample or population mean as an average or expected value, do not apply anymore and need to be replaced by barycenters or Fr\'echet means~\citep{frec:48}. In the regression context, the conditional Fr\'echet mean for random object response $Y$, residing in a metric space $(\Omega\lo Y,d\lo Y)$, given a Euclidean predictor $X\in \real \hi p$, is defined  as~\citep{hein:09,pete:19}
\begin{align}
	\label{cond:fr:mean}
	E\lo \oplus(Y|X=x) = m_\oplus(x) := \argmin_{y\in \Omega\lo Y} E[d\hi2\lo Y (Y,y)|X =x].
\end{align}
The Fr\'echet regression proposed by~\cite{pete:19} generalizes the globally linear least squares method
and the nonparametric local linear regression to fit the conditional Fr\'echet mean. They aim for direct modeling of the joint distribution of the response and the predictor by viewing the regression function as an alternative target of weighted Fr\'echet means, with weights that change globally linearly (or locally) with the predictors and are derived from those of the corresponding standard multiple linear regression (or local linear kernel regression) with Euclidean responses.
The globally linear approach, in particular, targets an alternative formulation than~\eqref{cond:fr:mean} given by
\begin{align}
	\label{global:linear:fr:mean}
	\tilde{m}_\oplus(x) = \argmin_{y\in \Omega\lo Y} E[s(X,x)d\hi2\lo Y (Y,y)],
\end{align}
where the weight function $s(X,x) = 1 + (x-\mu\lo X)\hi \top \Sigma\hi{-1}\lo X (X- \mu\lo X)$ varies globally and linearly with the output points $x\in \real \hi p$, hence the nomenclature;  $\mu\lo X$ and $\Sigma\lo X$ being the expectation and covariance matrix for the predictors $X$.

Model~\eqref{global:linear:fr:mean} coincides with model~\eqref{cond:fr:mean} in the special case of multiple linear regression with Euclidean responses and predictors.
However, for a general metric space-valued response $Y \in \Omega\lo Y,$ the above two targets are different, thus making the regression relationship for general metric-valued data quite restrictive.
Although the local regression, which indeed targets~\eqref{cond:fr:mean} with an asymptotically negligible bias, is more flexible, it is effective only when the dimension of the predictor is relatively low. As this dimension gets higher, its accuracy drops significantly\red{--}
a phenomenon known as the curse of dimensionality.
Recently~\cite{bhat:22} developed a single index Fr\'echet regression that projects the multivariate predictors onto a desired direction parameter vector to form a single index, thus facilitating inference for Fr\'echet regression. However, the model assumptions are still somewhat restrictive, and in general, the Fr\'echet regression framework can only accommodate Euclidean predictors.

In this work, we propose a non-linear global object regression framework that strikes a balance between the fully linear approach and the fully local approach. By mapping the predictor metric space into an RKHS, the new regression method offers the flexibility to accommodate a spectrum of model complexities such as the linear model, the polynomial model, and a family of functions that is dense in the $L \lo 2$ space. This flexibility is made possible via a novel probabilistic machinery that we call {\em the weak conditional Fr\'echet mean}, which is developed from the concept of weak conditional mean introduced by \cite{li:22} in the context of sufficient dimension reduction for functional data. It is important to note that there is no concept of linearity in an abstract metric space where the statistical objects reside--the model proposed in~\cite{pete:19} is called linear because of the linear form of the weight function through which the dependence of the response on the predictor is characterized in~\eqref{global:linear:fr:mean}. We develop the notion of a weak conditional Fr\'echet mean utilizing the smoothness in the predictor space and the intrinsic geometry implied by the metric in the response space, and introduce a novel nonlinear object regression approach as a generalization of nonlinear regression in metric spaces.

In addition to this flexibility, our method also allows both the response and the predictor to be metric-space-valued random objects. 
Studying the relation between two arbitrary random objects is also increasingly important. Unfortunately, not much exists in the literature in this regard, barring special cases of distribution-on-distribution regression~\citep{chen:19,chen:21,ghod:22}. Our proposed method accommodates more general predictors, such as random vectors, functions, or even object-valued predictors, as long as the predictor space admits an RKHS embedding. We discuss the details of constructing appropriate kernels to generate such  RKHSs and study the relevant operators generated to achieve this goal. Interestingly, in a special case, where the kernel for the RKHS is taken to be the linear kernel on a Euclidean space, our nonlinear global \frechet regression reduces to the (linear) global regression proposed by cite{pete:19}.

Along with--and also as a preparation for--our development of the nonlinear global \frechet regression,  we also give an in-depth development toward a coherent and comprehensive theoretical foundation for weak conditional mean and weak \frechet conditional mean, as we perceive they will play an increasingly important role in regression for functional data and metric-space-valued data. These serve as a bridge by which we can bring many tools available in classical regression to the new regression problems where the regression variables are random functions or random objects. In particular, we discuss the transparent and highly interpretable interrelations among four types of conditional means--the conditional mean, the weak conditional mean, the conditional \frechet mean, and the weak conditional \frechet mean  (see Figure~\ref{fig:cond:mean}).

The rest of the paper is organized as follows.  Section 2 defines the preliminary setup of the problem and focuses on the construction of the weak conditional mean for the classical/ Euclidean paradigm in detail. It is important to note that Section 2 by itself is a key contribution to the state-of-the-art literature for the  Hilbert space-valued functional data. Section 3 defines the weak condition moments for object responses and predictors, establishes the global non-linear object regression model, and studies its connections to the global linear object regression framework. In Section 4, we propose a suitable estimator for the weak conditional Fr\'echet mean from the observed data. In this vein, the construction of the underlying RKHS is discussed, and an M-estimation setting is devised. Section 5 establishes the asymptotic convergence rates of the proposed methods. Simulation results are presented in Section 6 to show the numerical performances of the proposed methods. Section 7 analyzes a real application of the proposed method for the mortality-vs-fertility distributions. All proofs are presented in Section S.1. of the Supplementary Material.

\section{Weak conditional mean and further development}

In this section, we first introduce the notations with a focus on the construction of a reproducing kernel Hilbert space on the space where the predictor objects lie. Next, we outline the basic idea underlying the construction of the weak conditional expectation in \cite{li:22}. We will also derive some new properties of weak conditional expectation and give a more general theory about the weak conditional expectation that is needed in later development.

\subsection{{\bf Random objects and reproducing kernels}}
Let $(\Omega, \ca F, P)$ be a probability space. Let $(\Omega \lo X, d \lo X)$ and $(\Omega \lo Y, d \lo Y)$ be metric spaces, where $\Omega \lo X$ and $\Omega \lo Y$ are set and $d \lo X$ and $d \lo Y$ are the metrics.  Let $\ca F \lo X$ and $\ca F \lo Y$ be the Borel $\sigma$-fields in $\Omega \lo X$ and $\Omega \lo Y$ corresponding to the open sets determined by $d \lo X$ and $d \lo Y$.  Let $X: \Omega \to \Omega \lo X$ and $Y: \Omega \to \Omega \lo Y$ be random elements that are measurable, respectively, with respect to $\ca F/ \ca F \lo X$ and $\ca F / \ca F \lo Y$.  Such random elements are called {\em statistical objects}. Let
$P \lo {XY} = P \of (X,Y)\inv$, $P \lo X = P \of X \inv$ and $P \lo Y = P \of Y \inv$ be the distributions of $(X,Y)$, $X$ and $Y$, respectively.

We will assume that there exists a positive definite kernel $\ka \lo X: \Omega \lo X \times \Omega \lo X \to \real$. While there are sufficient conditions for a  metric space to possess such kernels,  we make this requirement our general assumption.

\begin{assumption}\label{assumption:positive kernel} There is a positive definite kernel $\ka \lo X: \Omega \lo X \times \Omega \lo X \to \real$.
\end{assumption}

For example, if $\Omega \lo X$ is of negative type, then the metric-induced kernel is positive definite~\citep{sejd:13}. Furthermore,~\cite{zhan:21} showed that, if $\Omega \lo X$ is complete and separable, and there is a continuous injection from $\rho: \Omega \lo X \to \ca H$ for some separable Hilbert space $\ca H$, then, for any analytic function $F(t) = \sum \lo {i=1} \hi \infty a \lo i t \hi i$ with $a \lo i  > 0$, the function $\ka: \Omega \lo X \times \Omega \lo X \to \real$ of the form $F(\langle \rho (x \lo 1), \rho (x \lo 2) \rangle \lo {\ca H} )$ is a cc-universal kernel~\citep{micc:06}.

Let $\kappa_G(x,x') = \exp(-\gamma\lo X d\lo X \hi 2(x, x'))$ and $\kappa_L(x, x') = \exp(-\gamma \lo X d\lo X\hi 2(x, x'))$ denote the Gaussian and Laplacian kernels, respectively.~\cite{zhan:21} showed that both $\kappa_G$ and $\kappa_L$ on a complete and separable metric space $\Omega\lo X$ are positive definite and universal, and the RKHS $\ca M \lo X$ generated by such kernels is dense in $L^2(P_X)$.

Note that we do not impose the above assumption on $\Omega \lo Y$.

\subsection{{\bf Weak conditional mean via uncentered regression operator}}

We first define the extended Carleman operator, which is a slight extension of the definition in \cite{weid:12}.
\begin{definition} [Carleman operator]
	Let $\ca G$ be a set, $\ca M$ a Hilbert space of real-valued functions on $\ca G$, $\ca H$ another Hilbert space, and $A:\ca H \to \ca M$ a linear operator. If, for each $x\in \ca G$, the linear functional
	\[
	A\lo x: \ca H\rightarrow \real, \ f\mapsto (Af)(x)
	\]
	is bounded, then we call $A$ an extended Carleman operator. The Riesz representation $\lambda\lo A(x)$ of $A\lo x$ is called the inducing function of $A$.
\end{definition}
In the rest of the paper, $\ca G$ is the metric space $\Omega \lo X$, $\ca M \lo X$ is the RKHS generated by $\ka \lo X$, $\ca H$ is the real line $\real$, and $A: \real \to \ca M \lo X$ is the regression operator.

We next introduce the regression operator. Let $\ca H \lo U$ be a generic Hilbert space, and let $U: \Omega \to \ca H \lo U$ be a random element. We make the following assumption.
\begin{assumption}\label{assumption:separable} $\ca M \lo X$ and $\ca H \lo U$ are separable.
\end{assumption}
These conditions are mild: for example, by Theorem 2.7.5 of~\cite{hsin:15}, if $\Omega \lo X$ is separable and $\ka \lo X$ is continuous, then $\ca M \lo X$ is separable. Since $\ca H \lo U$ will be taken to be $\real$ for the rest of the paper, it is separable. Consider the tensor products
\begin{align*}
	\ka \lo X (\cdot, X) \otimes \ka \lo X (\cdot, X), \quad \ka \lo X (\cdot, X) \otimes U.
\end{align*}
The above quantities are members of the tensor product spaces $\ca M \lo X \otimes \ca M \lo X$ and $\ca M \lo X \otimes \ca H \lo U$, respectively. By simple calculation,
\begin{align}
	\begin{split}
		\ali \| \ka \lo X (\cdot, X) \otimes \ka \lo X (\cdot, X) \| \lo {\ca M \lo X \otimes \ca M \lo X} = \ka \lo X (X,X), \\
		\ali \| \ka  \lo X (\cdot, X) \otimes U \| \lo {\ca M \lo X \otimes \ca H \lo U}  = \sqrt{ \ka \lo X (X, X)} \| U \|.
	\end{split}\label{eq:norms}
\end{align}
We make the following assumption.
\begin{assumption}\label{assumption:finite moments} \quad $E \ka \lo X (X,X) < \infty$, $E ( \sqrt{ \ka \lo X (X, X)} \| U \|) < \infty$.
\end{assumption}
Since $\ca M \lo X$ and $\ca H \lo U$ are separable, $\ca M \lo X \otimes \ca M \lo X$ and $\ca M \lo X \otimes \ca H \lo U$ are separable. Furthermore, by Assumption \ref{assumption:finite moments} and relations in (\ref{eq:norms}), we have
\begin{align*}
	E ( \| \ka \lo X (\cdot, X) \otimes \ka \lo X (\cdot, X) \| \lo {\ca M \lo X \otimes \ca M \lo X} ) < \infty, \quad
	E ( \| \ka  \lo X (\cdot, X) \otimes U \| \lo {\ca M \lo X \otimes \ca H \lo U} ) < \infty.
\end{align*}
By Theorem 2.6.5 of~\cite{hsin:15}, the following Bochner integrals
\begin{align*}
	\int \lo \Omega  \ka \lo X (\cdot, X) \otimes \ka \lo X (\cdot, X) d P , \quad
	\int \lo \Omega  \ka  \lo X (\cdot, X) \otimes U  d P
\end{align*}
are defined. They will be denoted by $M \lo {XX}$ and $M \lo {XU}$, respectively, and will be called the covariance operator of $X$ and the cross-covariance operator from $\ca H \lo U$ to $\ca M \lo X$. It can be shown that, for any $f, g \in \ca M \lo X$ and $h \in \ca H \lo U$, we have
\begin{align}\label{eq:cov cov}
	\langle f, M \lo {XX} \rangle \lo {\ca M \lo X} = E [ f(X)g(X)], \quad \langle f, M \lo {XU} h \rangle \lo {\ca M \lo X} = E [f(X) \langle U, h \rangle \lo {\ca H \lo U}].
\end{align}

Henceforth, for a linear operator $A: \ca H \to \ca H$, let $\ran(A)$ denote the range of $A$ and $\ker (A)$ denote the kernel of $A$; that is, $\ran(A) = \{ Af: f \in \ca H\}$ and $\ker (A) = \{ f \in \ca H, A f  = 0 \}$.
Furthermore, let $\cran (A)$ denote the closure of $\ran(A)$. We make the following assumption.
\begin{assumption}\label{assumption:ker ran} \quad $\ker (M \lo {XX}) = \{0\}$ and $\ran (M \lo {XU} ) \subseteq \ran (M \lo {XX})$.
\end{assumption}
This assumption is very mild. By (\ref{eq:cov cov}), $M \lo {XX} f = 0$ implies $E[ f \hi 2 (X) ] = 0$, which implies that $f(X) = 0$ almost surely. If $\ka \lo X $ is continuous, then $f(X) = 0$ everywhere.  Hence, if $\ka \lo X$ is continuous, then $\ker (M \lo {XX} ) = \{ 0\}$. As argued in~\cite{li:18}, the assumption $\ran (M \lo {XU} ) \subseteq \ran (M \lo {XX})$ is  a smoothness assumption about the relation between $U$ and $X$. Under $\ker (M \lo {XX}) = \{0\}$, $M \lo {XX}: \ca M \lo X \to \ran( M \lo {XX})$ is an injective function. Thus the inverse function $M \lo {XX} \inv: \ran(M \lo {XX}) \to \ca M \lo X$ is defined. By $\ran (M\lo {XU} ) \subseteq \ran ( M \lo {XX})$, the operator
\begin{align*}
	R \lo {XU} = M \lo {XX} \inv M \lo {XU}
\end{align*}
is well-defined and is called the regression operator~\citep{lee:16}. Note, however, that since $M \lo {XX}$ is a trace class operator, $M \lo {XX} \inv$ is an unbounded operator. Nevertheless, as argued by~\cite{li:18}, it is entirely reasonable to assume $R \lo {XU}$ to be a bounded or even compact operator, which imposes a type of smoothness again on the relation between $U$ and $X$.
\begin{assumption}\label{assumption:bounded} \quad $R \lo {XU}: \ca H \lo U \to \ca M \lo X$ is a bounded operator.
\end{assumption}

As shown below, this assumption implies that $R \lo {XU}$ is an extended Carleman operator.
\begin{proposition}\label{proposition:bounded and carleman} If $R \lo {XU}$ is a bounded operator, then it is an extended Carleman operator.
\end{proposition}

The next theorem is the key property of the regression operator. Since it is more general than those given in~\cite{lee:16} and~\cite{li:22}, we provide a proof here.
\begin{theorem}\label{theorem:regression and conditional mean} If Assumptions \ref{assumption:positive kernel} through \ref{assumption:bounded} are satisfied and, for any $\alpha \in \ca H \lo U$, $E( \langle \alpha, Y \rangle \lo {\ca H \lo U} | X)$ is in the $L \lo 2 (P \lo X)$-closure of  $\ca M \lo X$, then
	\begin{enumerate}
		\item $E(\langle \alpha, Y \rangle \lo {\ca H \lo U} |X) \in \ran (R \lo {XU})$ almost surely;
		\item for any $\alpha \in \ca H \lo U$, $R \lo {XU}(\alpha )(X) = E [ \langle \alpha , U \rangle \lo {\ca H \lo U} | X ]$ almost surely.
	\end{enumerate}
\end{theorem}

As a special case, when $\ca M \lo {X}$ is dense in $L \lo 2(P \lo X)$, the conclusion of the theorem holds because in that case $E [ \langle \alpha, U \rangle \lo {\ca H \lo U} | X ]$ is always in the $L \lo 2 (P \lo X)$-closure of $\ca M \lo X$. This was the result proved in~\cite{li:22}.
The weak conditional mean is defined as the inducing function of the linear operator $R \lo {XU}$.
\begin{definition}\label{definition:regression operator} If Assumptions \ref{assumption:positive kernel} through \ref{assumption:bounded} are satisfied, then the random element
	\begin{align*}
		\omega \mapsto \lambda \lo {R \lo {XU}} (X (\omega)), \quad \Omega \to \ca H \lo U
	\end{align*}
	is the weak conditional expectation of $U$ given $X$; that is $\lambda \lo {R \lo {XU}} (X) = E ( U \bbar X)$.
\end{definition}

It follows easily from Theorem \ref{theorem:regression and conditional mean} that the weak conditional expectation reduces to the true conditional expectation under assumptions therein.

\begin{corollary}\label{corollary:weak and strong} Under the assumptions in Theorem \ref{theorem:regression and conditional mean}, we have
	\begin{align*}
		E ( U \bbar X) = E ( U |X).
	\end{align*}
\end{corollary}

\subsection{{\bf Weak conditional mean via centered regression operator}}

An alternative definition of the regression operator, as given in \cite{lee:16}, is the centered version of $R \lo {XU}$. Let
\begin{align*}
	\Sigma \lo {XX} = E [ ( \ka \lo X( \cdot, x) - \mu \lo X ) \otimes  ( \ka \lo X( \cdot, x) - \mu \lo X ) ], \quad
	\Sigma \lo {XU} = E [ ( \ka \lo X( \cdot, x) - \mu \lo X ) \otimes  ( U - \mu \lo U ) ].
\end{align*}
These operators are defined under Assumption \ref{assumption:finite moments}. We make a similar range assumption as Assumption \ref{assumption:ker ran}.

\begin{assumption}\label{assumption:ran ran centered} \quad $\ran (\Sigma \lo {XU}) \subseteq \ran (\Sigma \lo {XX})$.
\end{assumption}
In general, $\ker (\Sigma \lo {XX}) \ne \{0\}$, and so function $\Sigma \lo {XX}: \ca M \lo X \to \ca M \lo X$ is not invertible. However, the restricted operator
$\Sigma \lo {XX}|\lo{\cran (\Sigma \lo {XX})}$ is an invertible function. We call its inverse $[\Sigma \lo {XX}|\lo{\cran (\Sigma \lo {XX})}]\inv$ the Moore-Penrose inverse, and denote it by $\Sigma \lo {XX} \hi \dagger$. Note that this is a mapping from $\ran (\Sigma \lo {XX})$ to $\cran (\Sigma \lo {XX})$. Under Assumption \ref{assumption:ran ran centered},
the operator
\begin{align*}
	R \lo {XU} \hii c := \Sigma \lo {XX} \hi \dagger \Sigma \lo {XU}
\end{align*}
is well defined, and, to distinguish it from $R \lo {XU}$ above, we denote it by  $R \lo {XU} \hii c$ and call it the centered regression operator.

\begin{assumption}\label{assumption:bounded centered} \quad $R \lo {XU} \hii c$ is a bounded operator.
\end{assumption}

We now give the alternative definition of the weak conditional expectation using $R \lo {XU} \hii c$. It turns out that this alternative definition deals with the constant function %f(x)=\mbox{constant}$ 
better than the uncentered version.
\begin{definition}\label{definition:regression operator centered} Suppose $R _ {XU} \hii c$ is defined and is a Carleman operator. Then the following random element
	\begin{align*}
		E(U) + \lambda _ {R \lo {XU} \hii c}(X) - E[\lambda _ {R \lo {XU} \hii c}(X) ]
	\end{align*}
	is called the weak conditional expectation of $U$ given $X$ with respect to $\ca M \lo X$.
\end{definition}

The next proposition is a parallel result of Theorem \ref{theorem:regression and conditional mean} for the centered regression operator. We will say that a function $f$ belongs to a subset of $L \lo 2 (P \lo X)$ modulo constant if there is a constant $c$ such that $f+c$ belongs to that subset.

\begin{proposition}\label{proposition:regression and conditional mean} If Assumptions \ref{assumption:positive kernel}, \ref{assumption:separable},  \ref{assumption:finite moments}, \ref{assumption:ran ran centered}, and \ref{assumption:bounded centered} are satisfied and, for any $\alpha \in \ca H \lo U$, $E( \langle \alpha, Y \rangle \lo {\ca H \lo U} | X)$ belongs to  the $L \lo 2 (P \lo X)$-closure of  $\ca M \lo X$ modulo constant, then
	\begin{enumerate}
		\item $E(\langle \alpha, Y \rangle \lo {\ca H \lo U} |X) \in \ran (R \lo {XU})$ modulo constant almost surely;
		\item for any $\alpha \in \ca H \lo U$,
		\begin{align}\label{eq:parallel}
			E [ \langle \alpha , U \rangle \lo {\ca H \lo U} | X ]= \langle \alpha, E(U) \rangle \lo {\ca H \lo U} + R \lo {XU} \hii c (\alpha )(X) - E [R \lo {XU} \hii c (\alpha )(X)  ].
		\end{align}
	\end{enumerate}
\end{proposition}

The proof is similar to that of Theorem \ref{theorem:regression and conditional mean} and is omitted.
The advantage of Definition \ref{definition:regression operator centered} over Definition \ref{definition:regression operator} is that the former does not require the function $x \mapsto 1$ to be a member of $\ca M \lo X$, while the latter usually does, as shown in the next corollary. In the following, $\mathbbm{1}\lo X: \Omega \lo X \to \real$ stands for the function $x \mapsto 1$.

\begin{corollary}\label{corollary:one in M} Suppose
	\begin{enumerate}
		\item both $R \lo {XU}$ and $R \lo {XU} \hii c$ are defined and bounded;
		\item for any $\alpha \in \ca H \lo U$, $E ( \langle U, \alpha \rangle \lo {\ca H \lo U} |X)$ is in the $L \lo 2 (P \lo X)$-closure of $\ca M \lo X$;
		\item $E(U) - E [\lambda \lo  {R \lo {XU} \hii c} (X)]  \ne 0$.
	\end{enumerate}
	Then  $\mathbbm{1} \lo X$ belongs $\ca M \lo X$ almost surely.
\end{corollary}

The next simple example illustrates the advantage of $\mu \lo U + \lambda \lo {R \lo {XU} \hii c} (X) - E [\lambda \lo {R \lo {XU} \hii c} (X)]$ over $\lambda \lo {R \lo {XU}} (X)$ as the definition of weak conditional expectation.

\begin{example}\em Suppose $U$ and $X$ are random vectors in $\real \hi q$ and $\real \hi p$, respectively. Assume that
	\begin{align*}
		E(U|X) = a + B\trans X.
	\end{align*}
	where $a$ is a  nonzero vector in $\real \hi p$, and $B$ is a matrix in $\real \hi {p \times q}$. Under this model, it can be easily shown that
	\begin{align}\label{eq:EUX}
		E(U|X) = E(U) +  [\cov(U,X)][\var (X)]\inv (X- E(X)).
	\end{align}
	Let $\ca H \lo U$ be the Euclidean space $\real \hi q$ and $\ca M \lo X$ is the Hilbert space consisting of functions of the form $\{a \trans x: a \in \real \hi p \}$ with inner product defined by
	\begin{align*}
		\langle a \lo 1 \trans (\cdot), a \lo 2 \trans (\cdot) \rangle \lo {\ca M \lo X}= a \lo 1 \trans a \lo 2.
	\end{align*}
	The space $\ca M \lo X$ can be viewed as an RKHS with kernel $\ka \lo X ( a \lo 1 \trans (\cdot), a \lo 2 \trans (\cdot)) = a \lo 1 \trans a \lo 2$. In this case
	\begin{align*}
		M \lo {XX} = E [ ((\cdot) \trans X )\otimes ((\cdot) \trans X)], \quad M \lo {XU} = E [ ((\cdot) \trans X )\otimes U].
	\end{align*}
	The space $\ca M \lo X$ is isomorphic to $\real \hi p$ with the isomorphism $T: \ca M \lo X \to \real \hi p, \ a  \trans (\cdot) \mapsto a$. Furthermore,
	it can be easily shown that $
	T M \lo {XX} T \hi * = E (X X \trans)$ and $T M \lo {XU} = E (XU \trans)$.
	Hence
	\begin{align*}
		R \lo {XU}(\alpha)(X) =\ali  \langle R \lo {XU}(\alpha), (\cdot)\trans X \rangle \lo {\ca M \lo X} \\
		=\ali  (T R \lo {XU}(\alpha)) \trans    (T((\cdot)\trans X))   \\
		=\ali  (T R \lo {XU}(\alpha)) \trans    X  \\
		=\ali  (T M \lo {XX} \inv T \hi * T M \lo {XU} \alpha) \trans    X  \\
		=\ali \alpha \trans [(E(XX\trans)) \inv E(XU\trans) ] \trans    X,
	\end{align*}
	which implies
	$
	\lambda \lo {R \lo {XU}} =[(E(XX\trans)) \inv E(XU\trans) ] \trans    X.
	$
	Clearly, this is not the same as the right-hand side of (\ref{eq:EUX}).
	
	Next, let's consider the centered version. Similar to the above argument, we can show that
	\begin{align*}
		R \lo {XU} \hii c (\alpha)(X) =\ali \alpha \trans [(\var (X)) \inv \cov(X,U)] \trans    X,
	\end{align*}
	implying $\lambda \lo {R \lo {XU} \hii c } (X) = [(\var (X)) \inv \cov(X,U)] \trans    X$. Hence
	\begin{align*}
		E(U) + \lambda \lo {R \lo {XU} \hii c } (X) - E[\lambda \lo {R \lo {XU} \hii c } (X)]=E(U) +  [(\var (X)) \inv \cov(X,U)] \trans    (X-EX),
	\end{align*}
	which is exactly the right-hand side of (\ref{eq:EUX}). \eop
\end{example}
This example shows that when $\ca M \lo X$ does not contain $\onex$, $\lambda \lo {R \lo {XY}} (X)$ is not the right generalization of $E(U|X)$. In comparison,   $E(U) + \lambda \lo {R \lo {XU} \hii c } (X) - E[\lambda \lo {R \lo {XU} \hii c } (X)]$ gives the right generalization without requiring  $\ca M \lo X$ to contain $\onex$. The next theorem shows that when $\ca M \lo X$ does contain the $\onex$, the two definitions are equivalent.

\begin{theorem}\label{theorem:relation centered uncentered} If $R \lo {XU}$ and $R \lo {XU} \hii c$ are defined and bounded, and $\ca M \lo X$ contains $\onex$, then
	\begin{align*}
		\lambda \lo {R \lo {XU}} (X) = E (U) + \lambda \lo {R \lo {XU}\hii c} (X) - E[\lambda \lo {R \lo {XU}\hii c} (X) ]
	\end{align*}
	almost surely.
\end{theorem}

Throughout the rest of the paper, we will adopt Definition \ref{definition:regression operator centered} as our definition of the weak conditional expectation and denote it by $E(U \bbar X)$.
\section{Weak conditional Fr\'echet mean}	\label{sec:weak_fr_cond_mean}

\subsection{{\bf Weak conditional Fr\'echet mean and its properties}}
Having defined the weak conditional expectation of $E(U\bbar X)$, we now define the weak conditional \frechet mean of a random object $Y$ in the metric space $(\Omega \lo Y, d \lo Y)$. For any fixed $y \in \Omega \lo Y$, let $U = d \hi 2 (y, Y)$ and $\ca H \lo U = \real$. Assuming $(X, U)$ satisfies Assumptions Assumptions  \ref{assumption:positive kernel}, \ref{assumption:separable},  \ref{assumption:finite moments}, \ref{assumption:ran ran centered}, and \ref{assumption:bounded centered},  the weak conditional mean $E [ d \hi 2 ( y, Y) \bbar X ]$ is well defined.
\begin{definition} Suppose $X$  and $U = d \hi 2 (y, Y)$ satisfy Assumptions  \ref{assumption:positive kernel}, \ref{assumption:separable},  \ref{assumption:finite moments}, \ref{assumption:ran ran centered}, and \ref{assumption:bounded centered}. The weak conditional \frechet mean of $Y$ given $X$, denoted by $E \lo \oplus ( Y \bbar X = x)$, is the minimizer of $E [ d \hi 2 (y, Y) \bbar X = x]$. That is,
	\begin{align*}
		E \lo \oplus ( Y \bbar X = x) =  \argmin \lo {y \in \Omega \lo Y} E [ d\lo Y \hi 2 (Y, y) \bbar X = x].
	\end{align*}
	We use $E \lo \oplus (Y \bbar X)$ to denote the function $x\mapsto E \lo \oplus ( Y \bbar X = x)$.
\end{definition}
In plain language, the weak conditional \frechet mean is any minimizer (over $y \in \Omega \lo Y$) of the weak conditional mean of $d \hi 2 (y, Y)$ given $X$. The next proposition gives an explicit expression of $E(U \bbar X)$ when $U$ when $U$ is a random scalar.
\begin{corollary}\label{corollary:explicit weak mean} Suppose $\ca H \lo U = \real$ and $(X,U)$ satisfies Assumptions  \ref{assumption:positive kernel}, \ref{assumption:separable},  \ref{assumption:finite moments}, \ref{assumption:ran ran centered}, and \ref{assumption:bounded centered}. Then
	\begin{align}\label{eq:explicit}
		E (U \bbar X) = E(U) + \langle \ka \lo X (\cdot, X) - \mu \lo X, \Sigma \lo {XX} \hi \dagger E[(\ka \lo X (\cdot, X) - \mu \lo X )U ] \rangle \lo {\ca M \lo X}.
	\end{align}
	where $ (\ka \lo X (\cdot, x) - \mu \lo X )U$
	denotes the function $x \mapsto   (\ka \lo X (\cdot, x) - \mu \lo X )U.$
	%where $ E[(\ka \lo X (\cdot, X) - \mu \lo X )U ]$% \rangle \lo {\ca M \lo X}$
	% denotes the function $x \mapsto   E[(\ka \lo X (\cdot, x) - \mu \lo X )U ].$ % \rangle \lo {\ca M \lo X}$.
\end{corollary}

By this corollary, the weak condition \frechet mean can be written more explicitly as
\begin{align}	\label{eq:weak:cond:fr:mean:fina_form}
	\begin{split}
		f\lo\oplus(x) &:= E \lo \oplus (Y\bbar X=x)\\ &= \argmin \lo {y \in \Omega \lo Y} \bigl[
		E(d \hi 2 ( Y, y) ) + \ \langle \ka \lo X (\cdot, X) - \mu \lo X, \Sigma \lo {XX} \hi \dagger E[(\ka \lo X (\cdot, x) - \mu \lo X) d \hi 2 ( Y, y) ] \rangle \lo {\ca M \lo X} \bigr].
	\end{split}
\end{align}
Denoting $d\hi2\lo Y(Y,y)$ as $U(y)$, and the operator $E[(\ka \lo X (\cdot, X) - \mu \lo X) d \hi 2 ( Y, y)]$ as $\Sigma\lo{XU(y)}$ one can rewrite~\eqref{eq:weak:cond:fr:mean:fina_form} as
\begin{align}
	\label{eq:weak:cond:fr:mean:fina_form2}
	f\lo\oplus(x) = E \lo \oplus (Y\bbar X) = \argmin \lo {y \in \Omega \lo Y} \left[E(U(y)  ) + \langle \ka \lo X (\cdot, X) - \mu \lo X, \Sigma \lo {XX} \hi \dagger \Sigma\lo{XU(y)} \rangle \lo {\ca M \lo X} \right].
\end{align}

We take  $E \lo \oplus (Y\bbar X)$ as our population target for estimation in nonlinear global Fr\'echet regression, which offers great flexibility. First, when we employ a universal kernel such as the Gaussian kernel of the Laplacian kernel, we are guaranteed to recover the conditional \frechet mean. Indeed, by Proposition \ref{proposition:regression and conditional mean}, we have the following corollary.
\begin{corollary} Suppose $X$ and $U=d \lo Y (Y, y) \hi 2$ satisfy Assumptions   \ref{assumption:positive kernel}, \ref{assumption:separable},  \ref{assumption:finite moments}, \ref{assumption:ran ran centered}, and \ref{assumption:bounded centered}. If $\ca M \lo X$ is dense in $L \lo 2 (P \lo X)$ modulo constant, then
	\begin{align*}
		E \lo \oplus (Y|X) = E \lo \oplus (Y \bbar X).
	\end{align*}
\end{corollary}

Secondly, even when $\ca M \lo X$ is not dense in $L \lo 2 (P \lo X)$ modulo constant, it still makes sense to use $E \lo \oplus (Y \bbar X)$, because it has the following optimality property. Let $\ca N \lo X$ denote the  $L \lo 2 (P \lo X)$-closure of $\ca M \lo X + \spn (\onex)$. That is, a member of $\ca N \lo X$ can be written as the limit of functions of the form $f \lo n + c \lo n$, where $f \lo n \in \ca M \lo X$ and  $c \lo n$ is a constant.

\begin{theorem}\label{theorem:optimality} If $R \lo {XU} \hii c$ is defined and bounded, then, for any $f \in \ca N \lo X$,
	\begin{align*}
		E\{ [ E(U | X)  - E(U\bbar X) ] \hi 2 \} \le E\{ [ E(U | X)  -  f(X)] \hi 2 \}.
	\end{align*}
\end{theorem}

This theorem shows that even when $E \lo \oplus (Y \bbar X)$ is different from  $E \lo \oplus (Y | X)$,  the former is closest to the latter in the sense that the objective function by which we obtain the former is closer to the objective by which we obtain the latter than any other function in the $L \lo 2 (P \lo X)$-closure of $\ca M \lo X + \spn ( \onex)$.

When $\Omega \lo Y$ is a Hilbert space, say $\ca H \lo Y$, the weak \frechet conditional mean is defined as the minimizer of the weak conditional mean of the squared norm of the difference between $\| Y - y \| \hi 2 \lo {\ca H \lo Y}$. By making analogy with the fact that, in terms of the true conditional mean,  $E(Y|X)$ is indeed the minimizer of $E( \| Y = y \| \hi 2 | X)$, it seems plausible to expect that $E(Y\bbar X)$ is the minimizer of $E( \| Y - y \| \hi 2 \bbar X)$ over $\ca H \lo Y$. This is indeed the case, as shown in the next theorem.

\begin{theorem}\label{theorem:weak frechet c mean and weak c mean} If $\Omega \lo Y$ is a Hilbert space, $R \lo {UX}$ is defined and bounded, then
	\begin{align*}
		E \lo \oplus ( Y \bbar X) = E ( Y \bbar X).
	\end{align*}
\end{theorem}

\begin{figure}[!htb]
	\centering
	\includegraphics[width=.6\textwidth]{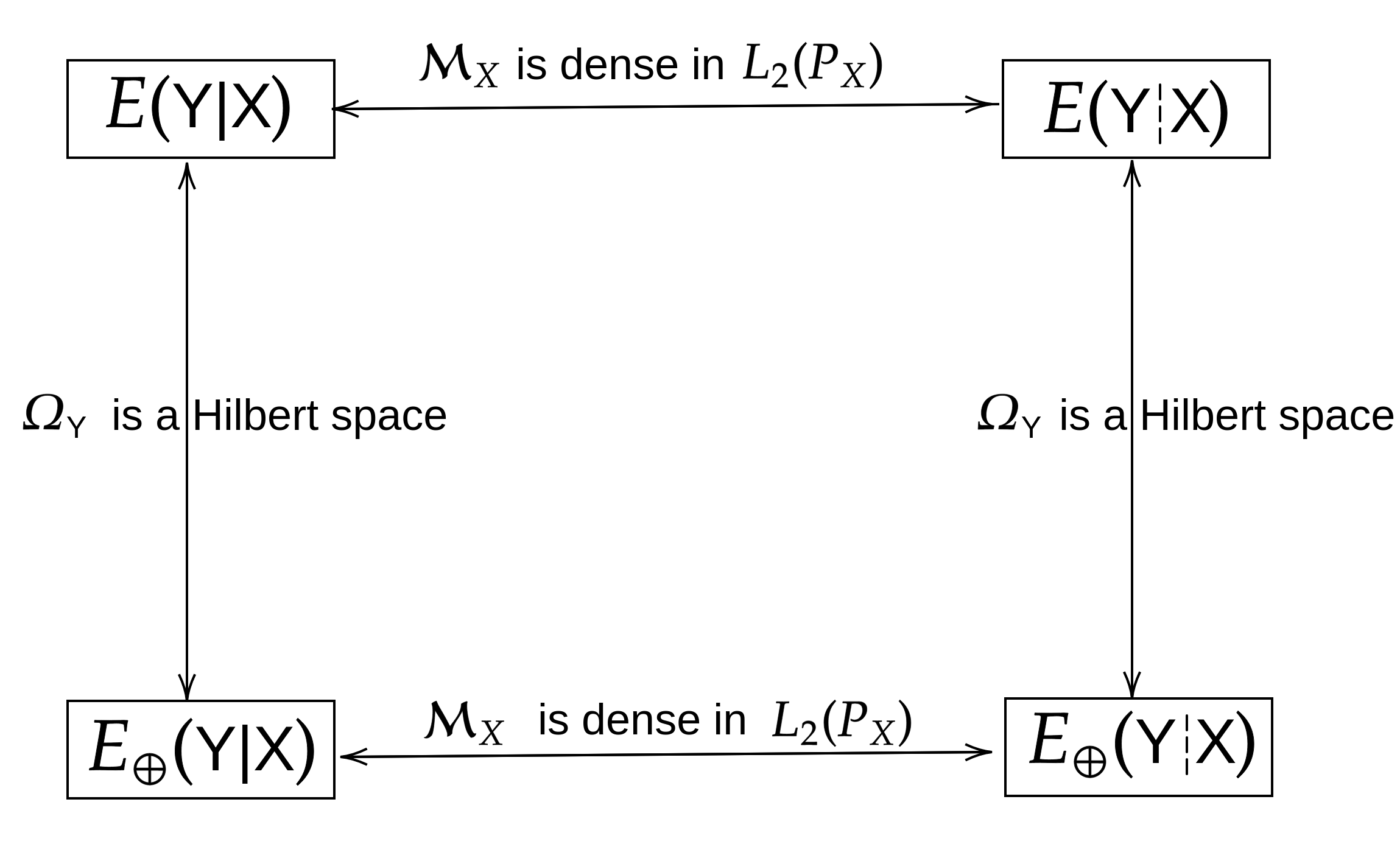}
	\centering
	\caption{Diagram describing the inter-relation between different types of conditional means.}
	\label{fig:cond:mean}
\end{figure}

So far, we have considered four types of conditional means: the conditional mean $E(Y|X)$, the \frechet conditional mean $E \lo \oplus (Y|X)$, the weak conditional mean $E(Y \bbar X)$, and the weak \frechet conditional mean $E \lo \oplus (Y \bbar X)$.
The conditional expectation $E(Y|X)$ can be seen as the orthogonal projection onto the closed subspace $L^2(P\lo X)$ that minimizes the expected squared difference $E(Y-X)^2$ among all random variables $X$, so in a sense, it is the best predictor of $Y$ based on the information in the  $\sigma$-algebra generated by a random variable $X$. Thus, more informally, $E(Y|X) = \Pi\lo{L\lo 2(P\lo X)}(Y).$ For random functions $X$ and $Y$ taking values in general Hilbert-spaces $\ca H\lo 1$ and $\ca H\lo 2$, respectively, weak conditional mean is given by the projection  $E(Y|X) = \Pi\lo{\ca H\lo 1}(Y).$ Both the concepts have now been generalized for metric space-valued data, and the next corollary summarizes their relations (also see Figure~\ref{fig:cond:mean}).

\begin{corollary}\label{corollary:summary} Suppose $R \lo {UX}$ is defined and bounded. Then
	\begin{enumerate}
		\item If $\Omega \lo Y$ is a Hilbert space, then
		\begin{align*}
			E \lo \oplus (Y|X) = E (Y|X), \quad E \lo \oplus (Y \bbar X) = E (Y \bbar X)
		\end{align*}
		\item If $\ca M \lo X$ is dense in $L \lo 2 (P \lo X)$ modulo constant, then
		\begin{align*}
			E ( Y | X) = E ( Y \bbar X), \quad E \lo \oplus ( Y | X) = E \lo \oplus ( Y \bbar X).
		\end{align*}
	\end{enumerate}
\end{corollary}

\subsection{{\bf Relation with global linear \frechet    regression}}

Interestingly, as the next theorem shows, the weak conditional \frechet means reduces to the objective function of the global linear \frechet regression introduced by ~\cite{pete:19} in a special case,  where $\ka \lo X$ is the linear kernel $c + x \lo 1 \trans x \lo 2$. Let $\Sigma \lo X = \var (X)$ be the covariance matrix of the random vector $X$.

\begin{theorem}
	\label{theorem:global linear frechet}
	If $\Sigma \lo X$ is invertible, $\ka \lo X$ is the linear kernel $c + x \lo 1 \trans x \lo 2$. Then
	\begin{align*}
		E[ d\lo Y \hi 2 (Y, y) \bbar X = x] = E \left \{ [ 1 + (x - EX) \trans \Sigma \lo X \inv (X - EX) ]  d\lo Y \hi 2 (Y, y) \right\}.
	\end{align*}
\end{theorem}

When $\ka\lo X$ is any arbitrary kernel such as a linear kernel and is not necessarily a universal kernel, the weak conditional Fr\'echet mean $E \lo \oplus (Y \bbar X)$ is not the same as the conditional Fr\'echet mean $E \lo \oplus (Y \bbar X)$. For example, as shown above, the target for the global Fr\'echet regression, which emerges as a special case of the weak conditional Fr\'echet means corresponding to a linear kernel, is different from the conditional Fr\'echet regression function $E \lo \oplus (Y|X)$. However, the regression relationship between two random objects $(X, Y)  \in \Omega\lo X \times \Omega\lo Y$ expressed through the weak Fr\'echet conditional mean is interesting and worth investigating in its own right. This alternative formulation is described through an RKHS embedding in the predictor space, thus accommodating random objects lying in the general metric space as a predictor. The characterization of the dependence between $Y$ and $X$ is global and nonlinear, and no bandwidth parameter is required to fine-tune the regression function.

\subsection{{\bf Existence and uniqueness of $E \lo \oplus (Y \bbar X)$}}
\label{sec:exist}

We now turn to the existence and uniqueness of the weak Fr\'echet conditional mean. Because the objective function $E \lo \oplus (d \hi 2 ( Y, y) | X)$ cannot, in general, be expressed as an integral with respect to a probability measure, the existing methods~\citep{afsa:11,  char:13, le:01, zeme:19} used for proving the existence and uniqueness for the Fr\'echet conditional mean cannot be used. Nevertheless, reasonably general statements about existence and uniqueness can be made under some conditions.

For existence,  by the extreme value theorem, if the function $y \mapsto E(d\lo Y\hi 2(Y, y)\bbar X=x)$ and $\Omega \lo Y$ is compact, then there is a $y \lo 0$ in $\Omega \lo Y$ that minimizes $E(d\lo Y\hi 2(Y, y)\bbar X=x)$, which then is a weak \frechet conditional mean.

We establish the existence and uniqueness of $E \lo \oplus (Y \bbar X)$ in two important special cases. The first case is where the metric space $\Omega \lo Y$ is of negative type, which guarantees that there is a continuous embedding from $\Omega \lo Y$ to a Hilbert space. 
\begin{definition}[Negative type metric space]
	The space $(M,\rho)$ with a semi-metric $\rho$ is of negative type if for all $n \geq 2$, $z_1, z_2, \dots,z_n \in M$ and $\alpha_1,\alpha_2,\dots,\alpha_n \in \real$, with $\sum_{i=1}^n \alpha_i = 0,$ one has $\sum_{i=1}^n\sum_{j=1}^n \alpha_i \alpha_j \rho(z_i,z_j) \leq 0$.
\end{definition}
The next theorem establishes the existence and uniqueness of $E \lo \oplus (Y \bbar X)$ rigorously when such an embedding exists. 
\begin{theorem}\label{proposition:unique negative type}
	Suppose Assumptions~\ref{assumption:positive kernel}-\ref{assumption:finite moments}, and \ref{assumption:ran ran centered}-\ref{assumption:bounded centered} are satisfied. Further, let the following conditions hold:
	\begin{enumerate}
		\item There is a Hilbert space  $\ca H$ and a continuous injection  $\rho: \Omega \lo Y \to \ca H$ such that  $\rho: \Omega \lo Y \to \rho (\Omega \lo Y)$ is an isometry.
		\item The set   $\rho(\Omega \lo Y)$ is convex and closed in $\ca H$.
	\end{enumerate}
	Then the  minimizer $E\lo{\oplus}(Y\bbar X) = \argmin \lo{ y \in \Omega\lo Y} E [ \|Y - y \|\lo {\ca H} \hi 2 \bbar X]$ exists and is unique.
\end{theorem}

The existence of such an isometric continuous map is not a strong requirement. For example, if $\Omega\lo Y$ is a separable metric space of negative type, one can always define the distance-induced kernel $\kappa:\Omega\lo Y\times \Omega\lo Y \to \real$ as
\[
\kappa(y,y') = \frac{1}{2}[d\lo Y(y, y_0) + d\lo Y(y',y_0) - d\lo Y(y, y')],
\]
for any fixed element $y_0 \in \Omega\lo Y.$ Then there us a unique RKHS $\ca H$ generated by this $\kappa$ and the map $\rho: \Omega\lo Y\to \ca H$ defined by $\rho(y) = \kappa(\cdot,y)$ satisfies all the requirements of the above proposition. Further, for many commonly observed object-valued data, the image set $\rho(\Omega\lo Y)$ is closed and convex in the underlying Hilbert space $\ca H$. Some examples are discussed in the following.\\

The second special case is where $\Omega \lo Y$ is a global nonpositive curvature metric space and  $\ca M \lo X$ is dense in $L \lo 2 (P \lo X)$ modulo constants. Again, let $U = d \lo Y(Y, y) \hi 2$.

\begin{proposition}\label{proposition:unique npc} Suppose
	\begin{enumerate}
		\item $R \lo {UY} \hii c$ is defined and bounded;
		\item $\ca M \lo X$ is dense in $L \lo 2 (P \lo X)$ modulo constants;
		\item $\Omega \lo Y$ is a global nonpositive curvature metric space.
	\end{enumerate}
	Then $E \lo \oplus ( Y \bbar X)$ exists and is unique.
\end{proposition}

For the definition and the related theories for a  global nonpositive curvature metric space, see \cite{stur:03}. The second special case is when $\Omega \lo Y$ is a negative-type metric space.  \\

\noindent \textit{Example 1:} The space of univariate probability distributions $G$ on $\real$ such that $\int_{\real}x^2 G(x)<\infty,$ equipped with the Wasserstein-2 metric. For two such distributions $G_1$ and $G_2$, the Wasserstein-2 metric between $G_1$ and $G_2$ is given by
\begin{align}
	\label{wass:dist}
	d\lo W \hi 2(G_1,G_2) = \int_0^1 (G\lo 1 \hi {-1}(t) - G\lo 2\hi {-1}(t))^2 dt,
\end{align}
where $G\lo 1 \hi {-1}$ and $G\lo 2 \hi {-1}$ are the quantile functions corresponding to $G_1$ and $G_2$, respectively.
The weak conditional Fr\'echet mean for distributional objects endowed with the Wasserstein-2 metric $d\lo W$ as defined above is given by the distributional object whose corresponding quantile function is equal to the $L^2([0,1])$-orthogonal projection of $E[Q\lo Y\bbar X]$ on $Q(\Omega\lo Y)$, where $Q(\Omega\lo Y)$ denotes the space of distributions represented as quantile functions and
\begin{align*}
	E[Q\lo Y\bbar X] = E(Q\lo Y) + \langle \ka\lo X(\cdot,x)-\mu\lo X,\ \Sigma\lo{XX}\hi \dagger \ E\left( (\ka\lo X(\cdot,X) - \mu\lo X)Q\lo Y\right)\rangle \lo {\ca M\lo X}.
\end{align*}

\noindent\textit{Example 2:}	The space of symmetric positive semi-definite matrices with unit diagonal, $\Omega \lo Y$, endowed with the Frobenius metric $d\lo F.$ For any two elements $A,B \in (\Omega \lo Y, d\lo F)$, their Frobenius distance is given by
\begin{align}
	\label{frob:dist}
	d\lo F\hi 2(A,B)= \sqrt{\text{trace }\left((A-B)(A-B)\hi \top\right)}.
\end{align}
The weak conditional Fr\'echet mean for spd matrix objects equipped with the Frobenius metric $d\lo F$ is given by the orthogonal projection of $B(x)$ onto the space of correlation matrices, where $B(x)$ has the $(j,k)$-th entry as
\begin{align*}
	B\lo{jk}(x) = E(Y\lo{jk}) + \langle \ka\lo X(\cdot,x)-\mu\lo X,\ \Sigma\lo{XX}\hi \dagger \ E\left( (\ka\lo X(\cdot,X) - \mu\lo X)Y\lo{jk}\right)\rangle \lo {\ca M\lo X}.
\end{align*}
Here $Y\lo{jk}$ is the $(j,k)$-th entry of $Y \in (\Omega\lo Y, d\lo F).$ The existence, uniqueness, and explicit form of the weak conditional Fr\'echet mean can also be derived for other Euclidean and pseudo-Euclidean metrics such as power metric, log-affine metric, Cholesky metric, etc.~\citep{dryd:10, lin:19}.
\section{Estimation}
\label{sec:theory}
In the last section, we have described the solution to the nonlinear object regression framework at the population level. In the following, we implement the regression at the sample level. The key steps involve the construction of the sample estimate for the regression function as an M-estimator based on \emph{i.i.d.} paired observations $(X_i, Y_i)_{i=1}^n$. In order to quantify the sample objective function minimized by the regression estimator, we need to express the underlying RKHS $\ca M \lo X$ and the relevant auto covariance and cross-covariance operators with a coordinate representation system (see, e.g.,~\cite{horn:12, li:18}).
\subsection{{\bf Coordinate representation}}
\label{sec:coord_rep}
Suppose that $\ca L \lo 1$ is a finite dimensional linear space with basis $\ca B=\{\xi\lo1, \xi\lo2,\dots, \xi\lo p\}.$ Then for any $\xi \in \ca L\lo 1,$ there is a unique vector $(a\lo1, a\lo2,\dots,a\lo p)\hi\intercal \in \real \hi p$ such that $\xi = \sum_{i=1}^{p}a_i\xi_i.$ The vector $(a\lo1, a\lo2,\dots,a\lo p)\hi\intercal$ is called the coordinate of $\xi$ with respect to $\ca B$, and denoted by $[\xi]\lo{\ca B}$. Throughout this section, we will use this notation to describe coordinate representation. Next, we introduce the coordinate representation of a linear operator between two (finite-dimensional) linear spaces. Suppose $\ca L \lo 2$ is another linear space with basis $\ca C = \{\eta_1,\eta_2,\dots,\eta_q\}$ and $A$ is a linear operator from $\ca L\lo1$ $\ca L\lo 2$. Then for any $\eta \in \ca L\lo 1$, we have
\begin{align*}
	A\xi &= A\left(\sum_{i=1}^p \left([\xi]\lo{\ca B}\right)\lo i \xi_i\right) = \sum_{i=1}^p \left([\xi]\lo{\ca B}\right)\lo i (A\xi_i) \\
	&= \sum_{i=1}^p \left([\xi]\lo{\ca B}\right)\lo i \sum_{j=1}^q \left([A\xi_i]\lo{\ca C}\right)_j \eta_j \ = \sum_{j=1}^q \left\{\left(\lo{\ca C}[A]\lo {\ca B}\right) \left([\xi]\lo {\ca B}\right) \right\}_j \eta_j,
\end{align*}
where $\lo {\ca C}[A]\lo {\ca B}$ is
the $q\times p$ matrix with $(i,j)$th entry $([A\xi_j]\lo{\ca C})_i$.
The above equation implies that $[A\xi]\lo{\ca C} = (\lo{\ca C}[A]\lo {\ca B})([\xi]\lo {\ca B}).$ Therefore we call the matrix $\lo{\ca C}[A]\lo {\ca B}$ the coordinate representation of the linear operator $A$ with respect to the bases $\ca B$ and $\ca C$. Similarly, for two Hilbert spaces $\ca H\lo 1$ and $\ca H\lo 2$, with spanning systems $\ca B \lo 1$ and $\ca B \lo 2$, and a linear operator $A: \ca H\lo 1 \to \ca H\lo 2$, we use the notation $\lo{\ca B \lo 1}[A]\lo {\ca B \lo 2}$ to represent the coordinate representation of $A$ relative to spanning systems $\ca B \lo 1$ and $\ca B \lo 2$.

\subsection{{\bf Construction of the RKHS $\ca M \lo X$ and model fitting}}
\label{sec:model:fit}
Let $(X_1,Y_1),\dots,(X_n,Y_n)$ be \emph{i.i.d.} observations of $(X,Y) \in \Omega\lo X\times \Omega\lo Y$. The RKHS $\ca M \lo X$ is spanned by $\{\ka \lo X(\cdot,X_i): i=1,\dots.n\}$ equipped with the inner product
\begin{align*}
	\langle f,g\rangle\lo{\ca M\lo X} = [f]\hi \top K\lo X [g],
\end{align*}
for any $f,g \in \ca M\lo X$, where $K\lo X$ is the $n\times n$ Gram matrix whose $(i,j)$th entry is $\ka \lo X(X_i,X_j)$, $i,j=1,\dots,n.$
Further, since the evaluation functional of the objective functions, the weak conditional Fr\'echet mean minimizes depend on $y\in \Omega\lo Y,$ we denote $U = U(y) = d\hi2\lo Y(Y,y).$ Similarly define $V(y) = d\lo Y(Y,y)$, and the sample observations as $U\lo i(y) = d\hi2\lo Y(Y\lo i,y)$ and $V\lo i(y) = d\lo Y(Y\lo i,y)$, respectively.

At the sample level, we estimate $\Sigma\lo{XX}$ $\Sigma\lo {XU(y)}$, and $\Sigma\lo{XV(y)}$ by replacing the expectations $E(\cdot)$ with the sample moments $E_n(\cdot)$ with respect to the empirical measure whenever possible. For example, we estimate $\Sigma\lo{XX}$ by $\hat{\Sigma}\lo{XX} = \frac{1}{n}\sum_{i=1}^{n} (\ka\lo X(\cdot,X_i) - \hat{\mu}\lo X) \otimes (\ka\lo X(\cdot,X_i) - \hat{\mu}\lo X),$ where $\hat{\mu}\lo X = \frac{1}{n}\sum_{i=1}^n \ka \lo X(\cdot,X_i).$ The sample estimates for $\Sigma\lo{XU(y)}$ and $\Sigma\lo{XV(y)}$, for any given $y \in \Omega\lo Y$, are similarly defined as $\hat{\Sigma}\lo{XU(y)} = \frac{1}{n}\sum_{i=1}^n (\ka\lo X(\cdot,X_i) - \hat{\mu}\lo X)U\lo i(y),$ and $\hat{\Sigma}\lo{XV(y)} = \frac{1}{n}\sum_{i=1}^n (\ka\lo X(\cdot,X_i) - \hat{\mu}\lo X)V\lo i(y),$ respectively.
Suppose, the subspace $\overline{\ran}(\hat{\Sigma}\lo{XX})$ is spanned by the set $\ca B \lo X = \left\{ \ka\lo X(\cdot,X_i) - E_n(\ka\lo X(\cdot,X_i)) : i =1,\dots,n\right\}.$
We then have the following coordinate representations of auto covariance and cross-covariance operators for any $y\in \Omega\lo Y$,
\begin{align*}
	\lo {\ca B\lo X} [\hat{\Sigma}\lo{XX}]\lo {\ca B \lo X}  = n\hi{-1}G_X,\ [\hat{\Sigma}\lo{XU(y)}]\lo{\ca B \lo X} =  [\hat{\Sigma}\lo{XV(y)}]\lo{\ca B \lo X} = n\hi{-1}G_X,\ \lo {\ca B\lo X} [\hat{\Sigma}\lo{XX}\hi \dagger]\lo {\ca B \lo X}  = n\hi{-1}G_X\hi \dagger,\
\end{align*}
where $G_X = QK_XQ$ and $G_X\hi \dagger$ is the Moore-Penrose inverse of $G_X$ via the Tikhonov-regularized inverse $(G_X +\epsilon_X I_n)\hi{-1}$ to prevent overfitting, where $\epsilon_X > 0$ is a tuning constant. Here $Q$ denotes the projection matrix $I_n - \frac{1}{n}1_n 1_n\hi \top$.
For a detailed discussion, see, for example, Section 12.4 of~\cite{li:18}.

Mimicking the definition of the population-level weak conditional Fr\'echet mean  $E\lo \oplus(Y\bbar X=x)$ from~\eqref{eq:weak:cond:fr:mean:fina_form} given by
\begin{align}
	\label{model}
	f\lo\oplus(x) &= \underset{y \in \Omega \lo Y}{\argmin\ } J(y),\text{ where }
	J(y) = E[U(y)] +   \langle \ka \lo X(\cdot, x) - \mu \lo X,  \Sigma \lo {XX} \hi \dagger \Sigma\lo{XU(y)} \rangle \lo {\ca M\lo X},
\end{align}
we define the following estimator
\begin{align}
	\label{estimator_final}
	\hat{f}\lo\oplus(x) &= \underset{y \in \Omega \lo Y}{\argmin\ } J_n(y),\text{ where }
	J_n(y) = \frac{1}{n}\sum_{i=1}^n U\lo i(y) +   \langle \ka \lo X(\cdot, x) - \hat{\mu} \lo X,  \hat{\Sigma} \lo {XX} \hi \dagger \hat{\Sigma}\lo{XU(y)} \rangle \lo {\ca M\lo X}.
\end{align}
To obtain a more explicit computable form of the above, it remains to identify the coordinate of $\ka \lo X (\cdot,x)-\hat{\mu}_X$ with respect to the spanning system $\{\ka \lo X(\cdot, X_i) -\hat{\mu}_X: i = 1,\dots,n\}.$ Suppose that $[\ka \lo X(\cdot,x)-\hat{\mu}_X] = c_x$ for some $c_x \in \real \hi n$. Then
\begin{align*}
	\langle \ka \lo X(\cdot,x)-\hat{\mu}_X, \ka \lo X(\cdot,X_i)-\hat{\mu}_X\rangle \lo {\ca M\lo X} = e_i\hi \top K_X c_x - \frac{1}{n} (e_i\hi \top K_X 1_n)(1_n\hi \top c_x) = e_i\hi \top K_X Q c_x,
\end{align*}
where $e_i$ denotes the vector whose $i\hi{\text{th}}$ component is $1$ and all others are $0$. Taking $i=1,\dots,n,$ we have $d\lo X  = K_XQc_x$, where $d\lo X $ is the vector of length $n$ with $i\hi{\text{th}}$ component $\ka \lo X(X_i,x) - E_n(\ka \lo X(X_i,x))$. With the Tikhonov regularization, we obtain the solution $c_x = Q(K_X +\epsilon_X I_n)\hi{-1}d\lo X $. Thus, the empirical objective function in~\eqref{estimator_final} becomes
\begin{align*}
	J_n(y) = \frac{1}{n} h_Y\hi \top 1_n + h_Y\hi \top G_X(G_X + \epsilon_X I_n)\hi{-1} c_x,
\end{align*}
where $h_Y$ is the vector with the $i\hi{\text{th}}$ component  $U\lo i(y)$, $i=1,\dots,n$, and $1_n = (1,1,\dots,1)\hi \top$.

\subsection{{\bf Tuning parameter selection}}
\label{sec:tuning}
We use the general cross-validation criterion~\citep{golu:79} to determine the tuning constant $\epsilon_X$ involved in the Tikhonov-regularization of the inverse auto-covariance operator $\Sigma\lo{XX}\hi\dagger$.
\begin{align}
	\label{gcv:tuning}
	\text{GCV}(\epsilon_X) = \frac{1}{n}\sum_{i=1}^n \frac{d\lo Y \hi 2(Y_i,\hat{Y}_i)}{\left(1 - \tr[QG_X(G_X +\epsilon_XI_n)\hi{-1} + 1_n 1_n\hi \top/n]/n\right)^2},
\end{align}
where $Y_i$ and $\hat{Y}_i$ are respectively the observed and predicted responses for the $i\hi{\text{th}}$ subject, $i=1,\dots,n.$
The numerator of this criterion quantifies the prediction error, while the denominator controls the degree of overfitting. We minimize the criterion over a grid $\{10^{-6},\dots,10\hi{-1}\}$ to find the optimal tuning constants.
\section{Convergence results}
\label{sec:conv}
In this section, we develop the asymptotic convergence results for the proposed Fr\'echet regression method. In particular, the convergence of the covariance operators with a suitable rate is established, which is used in turn to show the convergence of the regression estimate using the M-estimation theory.
\subsection{{\bf Convergence of regression operators}}
\label{sec:conv_reg_op}
The asymptotic properties of the empirical estimates of the mean and auto covariance operator defined on the RKHS $\ca M\lo X$ have been well-studied in the literature (see, for example, ~\cite{sang:22, fuku:07, lee:13, tao2022additive}). For completion, we list the properties here
\begin{lemma}
	\label{lem:conv_aux1}
	Under Assumptions \ref{assumption:positive kernel}-\ref{assumption:finite moments}, and \ref{assumption:ran ran centered}-\ref{assumption:bounded centered},
	%	and~\ref{ass:cov:smooth1}
	%		Under Assumptions \ref{assumption:positive kernel}, \ref{assumption:separable},  \ref{assumption:finite moments}, \ref{assumption:ran ran centered}, and \ref{assumption:bounded centered},
	%	and~\ref{ass:cov:smooth1}
	\begin{itemize}
		\item[(1)] $\lVert \hat{\mu}\lo X - \mu \lo X \rVert_{\ca M\lo X} = O_P(n\hi{-1/2}).$
		\item[(2)] $\lVert \hat{\Sigma}\lo{XX} - \Sigma \lo {XX}\rVert \lo {\text{OP}} = O_P(n\hi{-1/2})$.
	\end{itemize}
\end{lemma}

Suppose the eigenvalue and eigenfunction sequence of $\Sigma\lo{XX}$ is given by $\{(\lambda_j,\phi\lo j):\ j = 1,2,\dots \}$. By Mercer's theorem, the spectral decomposition of the auto covariance operator $\Sigma\lo {XX}$ is given by
\begin{align}
	\label{sigma:spectral}
	\Sigma\lo{XX} = \sum_{j=1}^\infty \lambda_j ( \phi\lo j \otimes\phi\lo j).
\end{align}
Typically, for a positive definite kernel $\ka\lo X$, $\Sigma\lo {XX}$ is a trace-class  operator whose eigenvalues decay to $0$, hence $\Sigma\lo {XX}\hi \dagger$ is unbounded. However, it is reasonable to assume the regression operators $	R\lo {XV(y)}:= \Sigma\lo {XX}\hi \dagger \Sigma\lo {XV(y)}$ and  $R\lo {XU(y)}:= \Sigma\lo {XX}\hi \dagger  \Sigma\lo {XU(y)}$ to be bounded uniformly for all $y\in \Omega\lo Y$.
We assume a degree of smoothness on the joint distribution of $(X, Y)$, requiring that the output functions for the regression operator must be sufficiently concentrated on the low-frequency components of $\Sigma\lo {XX}.$ %The following assumption is a stronger version of Assumption~\ref{assumption:bounded centered}.
\begin{assumption}
	\label{ass:cov:smooth1} $\underset{y\in \Omega\lo Y}{\sup \ }E\left((\phi\lo j(X) - E(\phi\lo j(X)))  \  d\lo Y\hi{k}(Y,y) \right) \leq \lambda_j^2$,\ $k = 1,2.$
\end{assumption}
The above condition implies that $R\lo {XU(y)} := \Sigma\lo{XX}\hi \dagger \Sigma\lo{XU(y)}$ and $R\lo {XV(y)} := \Sigma\lo{XX}\hi \dagger \Sigma\lo{XV(y)};$  are bounded operators uniformly for all $y \in (\Omega\lo Y,d\lo Y),$ in other words $\ran{(\Sigma\lo{XU(y)})}$, which can possibly depend on $y$, is entirely contained in the $\ran{(\Sigma\lo {XX})}$ uniformly across all possible $y \in \Omega\lo Y$, similarly for $\Sigma\lo{XV(y)}$. This is a generalization of Assumptions~\ref{assumption:ran ran centered} and~\ref{assumption:bounded centered} for the cross covariance operators indexed by $y\in \Omega\lo Y$ in the sense that the composite operators $\Sigma\lo{XX}\hi \dagger \Sigma\lo{XU(y)}$ and $\Sigma\lo{XX}\hi \dagger \Sigma\lo{XV(y)}$ are well-defined and bounded, uniformly for all $y \in \Omega\lo Y$. This can be interpreted as follows: $\Sigma\lo{XX}\hi \dagger \Sigma\lo{XU(y)}$ (and $\Sigma\lo{XX}\hi \dagger \Sigma\lo{XV(y)}$) must send all incoming functions into the low-frequency range of the eigenspaces of $\Sigma\lo{XX}$ with relatively large eigenvalues uniformly for all $y \in \Omega\lo Y$. That is, the  joint distribution of $(X,Y)$ is smooth enough such that the outputs of $\Sigma\lo{XU(y)}$ are the low-frequency components of $\Sigma\lo{XX}$, uniformly for all $y \in \Omega\lo Y$, similarly for $\Sigma\lo{XV(y)}$.

The consistent estimation for the cross-covariance operators is derived uniformly over all elements $y \in\Omega\lo Y,$ under the following assumption on the intrinsic geometry and complexity of the response space $(\Omega \lo Y, d\lo Y)$, which can be quantified by a bound on the entropy integral of $\Omega\lo Y.$
\begin{assumption}
	\label{ass:donsker}
	The entropy integral of $\Omega\lo Y$ is finite, i.e.,
	\[
	J := \int_{0}^{1} \sqrt{1 + \log N(\epsilon, \Omega \lo Y,d\lo Y)} d\epsilon <\infty,
	\]
	where $N(\epsilon, \Omega \lo Y,d)$ is the covering number for the space $\Omega\lo Y$ using balls of radius $\epsilon.$
\end{assumption}
This assumption is satisfied by most of the commonly observed random objects such as the space of univariate distributions with Wasserstein metric, space of positive semi-definite matrices with a suitable choice of metric, data on the surface of an $n-$ sphere with the intrinsic geodesic metric, and so on (see e.g.~\cite{dube:19} and the references therein).

\begin{proposition}
	\label{lem:conv_aux2}
	Under Assumptions~\ref{assumption:positive kernel}-\ref{assumption:finite moments}, and \ref{assumption:ran ran centered}-\ref{ass:donsker},
	%		Under Assumptions~\ref{assumption:positive kernel}, \ref{assumption:separable},  \ref{assumption:finite moments}, \ref{assumption:ran ran centered}, \ref{assumption:bounded centered},  \ref{ass:cov:smooth1}, and~\ref{ass:donsker},
	\begin{align*}
		\underset{y \in \Omega\lo Y}{\sup \ } \lVert\hat{\Sigma}\lo{XU(y)} - \Sigma\lo{XU(y)} \rVert \lo {\text{OP}} = O_P(n\hi{-1/2});\quad 	\underset{y \in \Omega\lo Y}{\sup \ } \lVert\hat{\Sigma}\lo{XV(y)} - \Sigma\lo{XV(y)} \rVert \lo {\text{OP}} = O_P(n\hi{-1/2}).
	\end{align*}
\end{proposition}

The consistent estimation for the regression operators is described in the following lemma under further smoothness conditions on the regression relationship between $X$ and $Y$.
\begin{assumption}
	\label{ass:rkhs:smooth2} For all $j\in \mathbb{N}$, there is a $0<\beta \leq 1$  such that\\ $\underset{y \in \Omega\lo Y}{\sup \ } E\left( (\phi\lo j(X) - E(\phi\lo j(X)))d\lo Y\hi k (Y,y)\right)\leq \lambda_j^{2+\beta}$, for $k=1,2,$, that is, there exists a bounded linear operator $S \lo {XY} : \ca M \lo X \to \ca M \lo X$ such that $\underset{y \in \Omega\lo Y}{\sup \ }\Sigma  \lo {XX} \hi {(1+\beta)\hi \dagger} \Sigma \lo{XU(y)}$ and $\underset{y \in \Omega\lo Y}{\sup \ }\Sigma  \lo {XX} \hi {(1+\beta)\hi \dagger} \Sigma \lo{XV(y)}$ are bounded linear operators uniformly over all $y \in \Omega \lo Y$.
\end{assumption}
Suppose $n\hi{-1/2} \prec \epsilon_n \prec 0$. For any $\beta$ as defined in Assumption~\ref{ass:rkhs:smooth2}, define
\begin{align}
	\label{eq:rkhs:rate}
	\alpha_n = \epsilon_n^\beta + \epsilon_n\hi{-1}n\hi{-1/2}.
\end{align}

\begin{proposition}
	\label{prop:conv_aux2}
	Under Assumptions~\ref{assumption:positive kernel}-\ref{assumption:finite moments}, and \ref{assumption:ran ran centered}-\ref{ass:rkhs:smooth2},
	%Under Assumptions~\ref{assumption:positive kernel}, \ref{assumption:separable},  \ref{assumption:finite moments}, \ref{assumption:ran ran centered}, \ref{assumption:bounded centered},  \ref{ass:cov:smooth1}, \ref{ass:donsker}, and~\ref{ass:rkhs:smooth2},
	\begin{align*}
		\underset{y \in \Omega\lo Y}{\sup \ }	\lVert 	\hat{\Sigma}\lo {XX} \hi \dagger \hat{\Sigma}\lo{XU(y)} - \Sigma \lo {XX} \hi \dagger \Sigma\lo{XU(y)}\rVert \lo {\text{OP}} &= O_P(\alpha_n), \\  	\underset{y \in \Omega\lo Y}{\sup \ }	\lVert 	\hat{\Sigma}\lo {XX} \hi \dagger \hat{\Sigma}\lo{XV(y)} - \Sigma \lo {XX} \hi \dagger\Sigma\lo{XV(y)}\rVert \lo {\text{OP}} &= O_P(\alpha_n),
	\end{align*}
	where $\alpha_n$ is as given in~\eqref{eq:rkhs:rate}.
\end{proposition}

\subsection{{\bf Estimation of weak conditional Fr\'echet mean}}
\label{sec:conv_weak_cond_fr_mean}
Having established the convergence of the regression operators, we proceed to derive the convergence results for the weak Fr\'echet conditional mean in~\eqref{estimator_final}. We require the following assumptions regarding the intrinsic geometry of the response space, which are the key to establishing the rate of convergence of any M-estimator, namely, the assumption of well-separateness of the minimizer, an upper bound on the entropy integral of the underlying metric space, and a local lower bound on the curvature of the objective functions listed in the Appendix.
\begin{theorem}
	\label{thm:probConv}
	Under Assumptions~\ref{assumption:positive kernel}-\ref{assumption:finite moments}, \ref{assumption:ran ran centered}-\ref{ass:rkhs:smooth2},  and the technical assumptions~\ref{ass:exists}-\ref{ass:entropy} in the Appendix, for any $x \in (\Omega \lo X, d\lo X )$,		
	%Under Assumptions~\ref{assumption:positive kernel}, \ref{assumption:separable},  \ref{assumption:finite moments}, \ref{assumption:ran ran centered}, \ref{assumption:bounded centered},  \ref{ass:cov:smooth1}, \ref{ass:donsker}, \ref{ass:rkhs:smooth2},  and the technical assumptions~\ref{ass:exists}, \ref{ass:entropy} in the Appendix, for any $x \in (\Omega \lo X, d\lo X )$,
	\begin{align*}
		d \lo Y(\hat{f}\lo\oplus(x), f\lo\oplus(x)) = o_P(1).
	\end{align*}
\end{theorem}

\begin{theorem}
	\label{thm:rate}
	Under Assumptions~\ref{assumption:positive kernel}-\ref{assumption:finite moments},~\ref{assumption:ran ran centered}-\ref{ass:rkhs:smooth2},  and the technical assumptions~\ref{ass:exists}-\ref{ass:curv} in the Appendix, with $\beta= 2$ in Assumption~\ref{ass:curv}, for any $x \in (\Omega \lo X, d\lo X )$,
	%Under Assumptions~\ref{assumption:positive kernel}, \ref{assumption:separable},  \ref{assumption:finite moments}, \ref{assumption:ran ran centered}, \ref{assumption:bounded centered},  \ref{ass:cov:smooth1}, \ref{ass:donsker}, \ref{ass:rkhs:smooth2},  and the technical assumptions~\ref{ass:exists}, \ref{ass:entropy}, and ~\ref{ass:curv} in the Appendix, for any $x \in (\Omega \lo X, d\lo X )$,
	\begin{align*}
		d \lo Y(\hat{f}\lo\oplus(x), f\lo\oplus(x)) = O_P(\alpha_n),
	\end{align*}
	where $\alpha_n$ is as given in~\eqref{eq:rkhs:rate}.
\end{theorem}	

For most commonly observed random objects $\beta$ in Assumption~\ref{ass:curv} is $2$, yielding an asymptotic rate of convergence for the M-estimator as $O_P(\alpha_n\hi{-1}).$ With a suitable rate from the RKHS regression literature, one can derive the rate of convergence as a function of the sample size $n$. For example, in ~\cite{li:17}, $\alpha_n \approx n^{-1/4}$, which is improved upon by~\cite{sang:22} as $\alpha_n \approx n^{-1/3}.$ This improved rate can be incorporated in the rate of convergence for the weak conditional Fr\'echet mean to yield an optimal rate of $O_P(n^{-1/3}).$

\section{Simulation studies}
\label{sec:simu}
In this section, we evaluate the numerical performances of the proposed nonlinear object-on-object regression method under different simulation settings for commonly observed random objects.

In all of the following simulation scenarios, we consider the Gaussian radial basis kernel $\kappa\lo G(y,y') = \exp(-\gamma \lo X d\hi 2(y, y'))$
as a candidate to construct the underlying RKHS $\ca M \lo X$ in the predictor space. We choose the parameters $\gamma\lo X$ as the fixed
quantities
\begin{align*}
	\gamma\lo X = \frac{\rho\lo Y}{2\sigma\lo G\hi 2},\ \sigma\lo G\hi 2 = \binom{n}{2}\hi{-1} \sum_{i<j} d\lo X\hi 2(X_i,X_j),\ \rho\lo Y = 1.
\end{align*}
The same choices of tuning parameters were used in~\cite{lee:13,li:17, zhan:22}. The metrics $d\lo X$ and $d\lo Y$ for the predictor and response metric spaces, respectively, are chosen appropriately to enhance the interpretability of the results in each of the following scenarios considered.
\paragraph*{{\bf Scenario 1: Univariate distribution-on-object regression}}
\label{sec:simu:univ:dist}
We consider univariate distributional objects as responses coupled with various types of statistical objects as predictors. Let $(\Omega\lo Y, d\lo Y)$ be the metric space of univariate distributions endowed with Wasserstein metric $d\lo Y = d\lo W$, as described in~\eqref{wass:dist} Section~\ref{sec:exist}. A sample of distributional object response, $Y_1,\dots, Y_n$ is taken in equivalent forms of either CDF, quantile functions, or densities. However, the distributions $Y_1,\dots, Y_n$ are usually not fully observed in practice, and the latent curves need to be recovered from the discrete observations $\{Y\lo {ij}\}_{j=1}^m$ for the $i\hi{\text{th}}$ sample; $i=1,\dots,n$, that one encounters in reality.  For this, we employ nonparametric smoothing with a suitable bandwidth choice implemented by the \emph{CreateDensity()} function in the \emph{frechet} R package~\citep{fr:package}. While  considering distributional predictors, the trajectories $X_i$ are recovered from the discrete observations $\{X\lo{ij}\}\lo{j=1}\hi m$; $i= 1\dots,n$ in a similar manner.

The random distributional response $Y$ is generated conditional on $X$ by adding noise to the quantile functions, which are demonstrated in the following simulation settings for various types of predictor objects. Generally, we let $Y = N (\zeta(x), \eta^2(x)),$ where the mean and variance of the response distribution are dependent on $X$. To this end, the auxiliary distribution parameters $\mu\lo Y$ and $\sigma\lo Y$, given $X$, are independently sampled such that $E(\mu\lo Y|X = x) = \zeta(x)$ and $E(\sigma\lo Y^2|X =x) = \eta^2(x)$, and the corresponding distributional response in its quantile representation is constructed as $Q\lo Y(\cdot) = \mu\lo Y + \sigma\lo Y \Phi\hi{-1}(\cdot)$.

To obtain the global nonlinear Fr\'echet regression estimator, one needs to solve the minimization problem in~\eqref{estimator_final}. We consider quantile function representation of the distributional responses. If $Q\lo{Y_i}$ is the quantile function corresponding to $Y_i$, $i =1,\dots,n;$ and $\hat{Q}\lo \oplus(\cdot;x)$ is the quantile function corresponding to the distribution $\hat{f}\lo\oplus(x)$ in~\eqref{estimator_final}, using similar logic as the proof of Proposition 4,
\begin{align*}
	\hat{Q}\lo \oplus(\cdot;x) = \argmin \lo{q \in Q(\Omega\lo Y)} \lVert q - \frac{1}{n}\sum\lo{i =1}\hi n w\lo{in}(x) Q\lo{Y_i}\rVert\lo{L^2[0,1]}.
\end{align*}
The existence and uniqueness of the solution of the above, and therefore of~\eqref{estimator_final}, is guaranteed $\hat{Q}\lo \oplus(\cdot;x)$ corresponds to the orthogonal projection of $g\lo x:= \frac{1}{n}\sum\lo{i =1}\hi n w\lo{in}(x) Q\lo{Y_i}$ as an element of the Hilbert space $L^2([0, 1])$ on the closed and convex set $Q(\Omega\lo Y)$, where $Q(\Omega\lo Y)$  is the space of quantile functions corresponding to distributions in $(\Omega\lo Y, d\lo W)$, as shown in Proposition 4. Here $w\lo{in} (x) = 1+ \langle \ka\lo X(\cdot,x)-\hat{\mu}\lo X, \hat{\Sigma}\lo{XX}\hi\dagger (\ka\lo X(\cdot, X\lo i)- \hat{\mu}\lo X )\rangle\lo{\ca M\lo X}$ is the nonlinear weight assigned to an observation at location $x$.

Taking an equidistant grid $\{u_j\}\lo{j=1}^M$ on $[0,1]$ and evaluating $g_j:= g\lo x(u_j)$, a discretized version, $\hat{Q}\hi \ast$, of the approximation of $\hat{Q}\lo \oplus(\cdot;x)$ is computed by solving the constrained quadratic program problem $\hat{Q}\hi \ast = \argmin\lo{q\in \real\hi M} \lVert g- q\rVert\lo E$ such that $q\lo 1\leq q\lo 2\dots\leq q\lo M.$ We employ an OSQP solver to implement this in practice.

We set the sample size $n = 200$ and $400$, and the number of discrete observations per sample $m = 50$ and  $100$ and generate the samples $(X_{i}, \{Y\lo {ij}\}_{j=1}^m)_{i=1}^n$. We use half of the samples to train the predictors via the proposed object regression method and then evaluate the prediction error as the discrepancy between the estimated and true responses using the rest of the data set by computing the Wasserstein distance metric~\eqref{wass:dist} between the two distributions. The tuning parameter for the Tikonov regularization is determined by the method described in Section~\ref{sec:tuning}. The experiment is repeated $B = 100$ times, and averages of the prediction error are computed as
\begin{align}
	\label{simu:rmpe1}
	\text{MPE} := \frac{1}{B} \sum_{b=1}^B d\lo W (Y\hi{\text{test}}\lo b, \hat{Y}\hi{\text{test}}\lo b),
\end{align}
where $Y\hi{\text{test}}\lo b$ and $\hat{Y}\hi{\text{test}}\lo b$ are the observed and predicted responses in the test set, respectively, for the $b$-th replicate, $b= 1\dots, B.$ The standard errors are also computed and will be reported in parentheses. \\

\noindent \textbf{Model I.1 (Euclidean predictors)}:
$\mu\lo Y|X \sim N((\beta\hi \top X)^2,\nu \lo 1 \hi 2)$  and $\sigma\lo Y|X \sim \\$ $Gamma((\gamma\hi \top X)^2/\nu \lo 2, \nu \lo 2/(\gamma\hi \top X)).$\\

\noindent \textbf{Model I.2 (Euclidean predictors)}:
After sampling the distribution parameters as in the previous setting, the resulting distribution is then ``transported'' in Wasserstein space via a random transport map $T$, that is uniformly sampled from a family of perturbation/ distortion functions $\{T_k: k \in \pm1, \pm 2,\}$, where $T_k(x) =  x -\sin(kx)/|k|$. The transported distribution is given by $T\#(\mu\lo Y + \sigma\lo Y \Phi\hi{-1})$, where $T\#p$
is a push-forward measure such that $T\#p(A) = p\left( \{ x: T(x) \in A\}\right)$, for any measurable function $T:\real \to \real$, distribution $p\in (\Omega\lo Y,d\lo W)$, and set $A \subset \real$. We sample the random transport map $T$ uniformly
from the collection of maps described above; $p$ denotes a Gaussian distribution with
parameters $\zeta(x) = (\beta\hi \top X)\hi2$ and $\eta\hi 2(x) = (\gamma\hi \top X)\hi2$.
The distributions thus generated are not Gaussian anymore due to transportation. The conditional Fr\'echet mean can be shown to remain at $\mu\lo Y + \sigma\lo Y \Phi\hi{-1}$ as before.\\

For Models I.1 and I.2, the Euclidean vector predictor $X \in \real \hi p$ is generated as follows: (i) we first generate $U_1,\dots, U_p$ from the AR(1) model with mean $0$ and covariance matrix $\Sigma = (0.5\hi {|i-j|})_{i,j}$, and then (ii) generate $X_j = 2\Phi(U_j) -1$, $j=1,\dots,p,$ where $\Phi$ is the c.d.f. of $N(0, 1)$. We select
$\nu\lo 1\hi2 = 0.1$, $\nu \lo 2 = 0.25$, $\beta = (1,-2,0,1)$, and $\gamma = (0.1,0.2,1,0.3)\hi \top$ in the above models.

The performance of our method, denoted by global nonlinear Fr\'echet regression (GNLFR), is compared with the globally linear Fr\'echet regression (GLFR) method by~\cite{pete:19}, which can only accommodate vector-valued predictors. We compute the MPE in~\eqref{simu:rmpe1} for varying levels of the predictor dimension, sample size, and number of discrete observations for each sample of distributions, namely $p,n,$ and $m$, respectively. Table~\ref{tab:sim:sceI1} summarizes the results. The prediction error decreases generally corresponding to a lower dimension $p$ of the predictor, a larger sample size $n$, and a denser design (higher $m$) over which the response is sampled. Across the board, our method outperforms the GLFR method regarding prediction accuracy. In setting I.1, when the underlying model is more linear, which is the ideal setting for the GLFR method, our method (GNLFR) has a competitive performance. Further, for setting I.2 the GNLFR method proves significantly better, which is not unexpected given the highly non-linear data-generating mechanism for this setting.
\begin{table}[!htb]
	%\centering
	\caption{Performances of the proposed global nonlinear Fr\'echet regression (GNLFR) and the global linear Fr\'echet regression by~\cite{pete:19} (GLFR) for univariate distributional responses with Euclidean predictors under Models I.1-I.2. The lowest number in a row corresponding to each data-generating mechanism is highlighted.}
	\label{tab:sim:sceI1}
	\centering
	\begin{tabular}{|c||cc|cc||cc|cc|}
		\hline
		& \multicolumn{2}{c|}{I.1 (GNLFR)} & \multicolumn{2}{c||}{I.1 (GLFR)} & \multicolumn{2}{c|}{I.2 (GNLFR)} &
		\multicolumn{2}{c|}{I.2 (GLFR)} \\ \hline
		($p$,$n$)\textbackslash{}$m$ & \multicolumn{1}{c|}{50} & 100 & \multicolumn{1}{c|}{50} & 100 & \multicolumn{1}{c|}{50} & 100 &
		\multicolumn{1}{c|}{50} & 100 \\ \hline
		(4,200) & \multicolumn{1}{c|}{\begin{tabular}[c]{@{}c@{}}0.037 \\ (0.012)\end{tabular}} & \begin{tabular}[c]{@{}c@{}} 0.024 \\ (0.016)\end{tabular} & \multicolumn{1}{c|}{\begin{tabular}[c]{@{}c@{}}0.033 \\ (0.021)\end{tabular}} & \begin{tabular}[c]{@{}c@{}} \textbf{0.018} \\ (0.014)\end{tabular} & \multicolumn{1}{c|}{\begin{tabular}[c]{@{}c@{}}0.110 \\ (0.081)\end{tabular}} & \begin{tabular}[c]{@{}c@{}} \textbf{0.087} \\ (0.070)\end{tabular} &
		\multicolumn{1}{c|}{\begin{tabular}[c]{@{}c@{}}0.230 \\ (0.012)\end{tabular}} &
		\begin{tabular}[c]{@{}c@{}} 0.181 \\ (0.011)\end{tabular} \\ \hline
		(10,200) & \multicolumn{1}{c|}{\begin{tabular}[c]{@{}c@{}}0.051 \\ (0.019)\end{tabular}} & \begin{tabular}[c]{@{}c@{}} 0.042 \\ (0.015)\end{tabular} & \multicolumn{1}{c|}{\begin{tabular}[c]{@{}c@{}}0.054 \\ (0.017)\end{tabular}} & \begin{tabular}[c]{@{}c@{}} \textbf{0.039} \\ (0.020)\end{tabular} & \multicolumn{1}{c|}{\begin{tabular}[c]{@{}c@{}}0.187 \\ (0.031)\end{tabular}} & \begin{tabular}[c]{@{}c@{}} \textbf{0.112} \\ (0.023)\end{tabular} &
		\multicolumn{1}{c|}{\begin{tabular}[c]{@{}c@{}}0.334 \\ (0.045)\end{tabular}} &
		\begin{tabular}[c]{@{}c@{}} 0.278 \\ (0.031)\end{tabular} \\ \hline
		(20,200) & \multicolumn{1}{c|}{\begin{tabular}[c]{@{}c@{}}0.058 \\ (0.018)\end{tabular}} & \begin{tabular}[c]{@{}c@{}} 0.051 \\ (0.018)\end{tabular} & \multicolumn{1}{c|}{\begin{tabular}[c]{@{}c@{}}0.061 \\ (0.020)\end{tabular}} & \begin{tabular}[c]{@{}c@{}}\textbf{0.045} \\ (0.019)\end{tabular} & \multicolumn{1}{c|}{\begin{tabular}[c]{@{}c@{}}0.210 \\ (0.029)\end{tabular}} & \begin{tabular}[c]{@{}c@{}} \textbf{0.153} \\ (0.028)\end{tabular} &
		\multicolumn{1}{c|}{\begin{tabular}[c]{@{}c@{}}0.431 \\ (0.025)\end{tabular}} &
		\begin{tabular}[c]{@{}c@{}} 0.391\\ (0.022)\end{tabular}\\ \hline
		(4,400) & \multicolumn{1}{c|}{\begin{tabular}[c]{@{}c@{}}0.021 \\ (0.009)\end{tabular}} & \begin{tabular}[c]{@{}c@{}} \textbf{0.013} \\ (0.009)\end{tabular} & \multicolumn{1}{c|}{\begin{tabular}[c]{@{}c@{}}0.034 \\ (0.010)\end{tabular}} & \begin{tabular}[c]{@{}c@{}} 0.021\\ (0.011)\end{tabular} & \multicolumn{1}{c|}{\begin{tabular}[c]{@{}c@{}}0.089 \\ (0.021)\end{tabular}} & \begin{tabular}[c]{@{}c@{}} \textbf{0.047} \\ (0.022)\end{tabular} &
		\multicolumn{1}{c|}{\begin{tabular}[c]{@{}c@{}}0.134 \\ (0.020)\end{tabular}} &
		\begin{tabular}[c]{@{}c@{}}0.086 \\ (0.021)\end{tabular}\\ \hline
		(10,400) & \multicolumn{1}{c|}{\begin{tabular}[c]{@{}c@{}}0.029 \\ (0.010)\end{tabular}} & \begin{tabular}[c]{@{}c@{}} 0.024 \\ (0.011)\end{tabular} & \multicolumn{1}{c|}{\begin{tabular}[c]{@{}c@{}}0.037 \\ (0.009)\end{tabular}} & \begin{tabular}[c]{@{}c@{}} \textbf{0.023} \\ (0.008)\end{tabular} & \multicolumn{1}{c|}{\begin{tabular}[c]{@{}c@{}}0.174 \\ (0.019)\end{tabular}} & \begin{tabular}[c]{@{}c@{}} \textbf{0.133} \\ (0.020)\end{tabular} &
		\multicolumn{1}{c|}{\begin{tabular}[c]{@{}c@{}}0.356 \\ (0.012)\end{tabular}} &
		\begin{tabular}[c]{@{}c@{}}0.239 \\ (0.014)\end{tabular}\\ \hline
		(20,400) & \multicolumn{1}{c|}{\begin{tabular}[c]{@{}c@{}}0.041 \\ (0.013)\end{tabular}} & \begin{tabular}[c]{@{}c@{}} \textbf{0.033} \\ (0.011)\end{tabular} & \multicolumn{1}{c|}{\begin{tabular}[c]{@{}c@{}}0.081 \\ (0.015)\end{tabular}} & \begin{tabular}[c]{@{}c@{}}0.043 \\ (0.015)\end{tabular} & \multicolumn{1}{c|}{\begin{tabular}[c]{@{}c@{}}0.189 \\ (0.016)\end{tabular}} & \begin{tabular}[c]{@{}c@{}} \textbf{0.122} \\ (0.016)\end{tabular} &
		\multicolumn{1}{c|}{\begin{tabular}[c]{@{}c@{}}0.451 \\ (0.013)\end{tabular}} &
		\begin{tabular}[c]{@{}c@{}}0.378 \\ (0.015)\end{tabular}\\ \hline
	\end{tabular}
\end{table}

For Models I.3-I.5 below, we consider univariate distribution-on-distribution regression.\\

\noindent \textbf{Model I.3 (Univariate distributions as predictors)}: $\mu\lo Y|X \sim N( \exp(W_2^2(X,\mu \lo 1)) +\\$ $\exp(W_2^2(X,\mu \lo 2)),\nu \lo 1 \hi 2)$ and $\sigma\lo Y|X = 0.1.$\\

\noindent \textbf{Model I.4 (Univariate distributions as predictors)}:  $\mu\lo Y|X \sim N( \exp(W_2^2(X,\mu \lo 1))$\\ $,\nu \lo 1 \hi 2)$ and $\sigma\lo Y|X = Gamma(W_2^2(X,\mu \lo 2), W_2(X,\mu \lo 2)).$\\

\noindent \textbf{Model I.5 (Univariate distributions as predictors)}:  $\mu\lo Y|X \sim N(\exp(H(X, \mu \lo 1)),$\\ $0.2^2); \sigma\lo Y|X = \exp(H(X, \mu \lo 2))$.\\

In the above we let $\nu \lo 1 \hi 2 = 0.1$, $\mu \lo 1 = Beta(2, 1)$ and $\mu \lo 2 = Beta(2, 3)$ and generate discrete observations from distributional predictors by $\{X\lo {ij}\}_{j=1}^m \overset{i.i.d.}{\sim} Beta(a_i, b_i)$, where $a_i \overset{i.i.d.}{\sim} Gamma(2, \text{rate} = 1)$ and
$b_i \overset{i.i.d.}{\sim} Gamma(2, \text{rate} = 3).$ $W_2(\cdot,\cdot)$ and $H(\cdot,\cdot)$ denote, respectively, the Wasserstein-2 distance and the Hellinger distance between two univariate distributional objects. The Hellinger distance between two Beta distributions $\mu = Beta(a_1, b_1)$ and $\nu = Beta(a_2, b_2)$ can be represented explicitly as
\begin{align*}
	H(\mu,\nu) &= 1 - \int\sqrt{f_\mu(t)f_\nu(t)} dt = 1- \frac{B((a_1 + a_2)/2, (b_1 + b_2)/2)}{\sqrt{B(a_1,b_1) B(a_2,b_2)}},
\end{align*}
where $B(\alpha, \beta)$ is the $Beta$ function.\\

Note that by virtue of the Gram matrix of the underlying RKHS kernel $\ka\lo X$, the predictor space is now embedded into a Hilbert space, hence finding the weak conditional Fr\'echet mean reduces to solving a constrained quasi-quadratic optimization problem and projecting back into the solution space.

The performance of our method, denoted by global nonlinear Fr\'echet regression (GNLFR), is compared with the distribution-on-distribution Wasserstein regression (WR) proposed by~\cite{chen:21} for varying choices of the sample size and predictor dimension $(n,m)$ (see Table~\ref{tab:sim:sceI2}).
\begin{table}[!htb]
	\centering
	\caption{Performances of the proposed global nonlinear Fr\'echet regression (GNLFR) and the Wasserstein Regression (WR) method by~\cite{chen:21} for univariate distribution-on-distribution regression under Models I.3- I.5. The lowest number in a row corresponding to each data-generating mechanism is highlighted.}
	\centering
	\label{tab:sim:sceI2}
	\begin{tabular}{|c||cl|cl|cl|c||c|c|}
		\hline
		($n$, $m$) & \multicolumn{2}{c|}{\begin{tabular}[c]{@{}c@{}}I.3 \\ (GNLFR)\end{tabular}} & \multicolumn{2}{c||}{\begin{tabular}[c]{@{}c@{}}I.3 \\ (WR)\end{tabular}} & \multicolumn{2}{c|}{\begin{tabular}[c]{@{}c@{}}I. 4 \\ (GNLFR)\end{tabular}} & \begin{tabular}[c]{@{}c@{}}I.4 \\ (WR)\end{tabular} & \begin{tabular}[c|]{@{}c@{}}I.5 \\ (GNLFR)\end{tabular} & \begin{tabular}[c]{@{}c@{}}I.5 \\ (WR)\end{tabular} \\ \hline
		(200, 50) & \multicolumn{2}{c|}{\begin{tabular}[c]{@{}c@{}}0.314\\ (0.121)\end{tabular}} & \multicolumn{2}{c||}{\begin{tabular}[c]{@{}c@{}}\textbf{0.298}\\ (0.191)\end{tabular}} & \multicolumn{2}{c|}{\begin{tabular}[c]{@{}c@{}}\textbf{0.461}\\ (0.110)\end{tabular}} & \begin{tabular}[c]{@{}c@{}} 0.514\\ (0.093)\end{tabular} & \begin{tabular}[c]{@{}c@{}}\textbf{0.491}\\ (0.110)\end{tabular} & \begin{tabular}[c]{@{}c@{}}0.820\\ (0.217)\end{tabular} \\ \hline
		(200, 100) & \multicolumn{2}{c|}{\begin{tabular}[c]{@{}c@{}} \textbf{0.268}\\ (0.091)\end{tabular}} & \multicolumn{2}{c||}{\begin{tabular}[c]{@{}c@{}}0.272\\ (0.110)\end{tabular}} & \multicolumn{2}{c|}{\begin{tabular}[c]{@{}c@{}}\textbf{0.381}\\ (0.125)\end{tabular}} & \begin{tabular}[c]{@{}c@{}}0.443\\ (0.112)\end{tabular} & \begin{tabular}[c]{@{}c@{}}\textbf{0.407}\\ (0.099)\end{tabular} & \begin{tabular}[c]{@{}c@{}}0.788\\ (0.098)\end{tabular} \\ \hline
		(400, 50) & \multicolumn{2}{c|}{\begin{tabular}[c]{@{}c@{}}0.159\\ (0.092)\end{tabular}} & \multicolumn{2}{c||}{\begin{tabular}[c]{@{}c@{}}\textbf{0.155}\\ (0.082)\end{tabular}} & \multicolumn{2}{c|}{\begin{tabular}[c]{@{}c@{}}\textbf{0.218}\\ (0.160)\end{tabular}} & \begin{tabular}[c]{@{}c@{}}0.310\\ (0.188)\end{tabular} & \begin{tabular}[c]{@{}c@{}}\textbf{0.251}\\ (0.181)\end{tabular} & \begin{tabular}[c]{@{}c@{}}0.549\\ (0.167)\end{tabular} \\ \hline
		(400, 100) & \multicolumn{2}{c|}{\begin{tabular}[c]{@{}c@{}}\textbf{0.134}\\ (0.086)\end{tabular}} & \multicolumn{2}{c||}{\begin{tabular}[c]{@{}c@{}}0.141\\ (0.079)\end{tabular}} & \multicolumn{2}{c|}{\begin{tabular}[c]{@{}c@{}}\textbf{0.172}\\ (0.155)\end{tabular}} & \begin{tabular}[c|]{@{}c@{}}0.256\\ (0.167)\end{tabular} & \begin{tabular}[c]{@{}c@{}}\textbf{0.177}\\ (0.120)\end{tabular} & \begin{tabular}[c]{@{}c@{}}0.422\\ (0.115)\end{tabular} \\ \hline
	\end{tabular}
\end{table}
We observed a decrease in the MPE as per~\eqref{simu:rmpe1} for all the settings as the sample size $n$ was increased favorably for the denser design with a higher $m$. For setting I.3, our method fairs comparably well with the WR method, but for more non-linear data generation mechanisms, as in settings I.4 and I.5, our method outperforms the WR method. Further, our method uses the intrinsic geometry of the space, as compared to the WR method, which utilizes the pseudo-Riemannian structure of the Wasserstein space, thus making our estimation more reliable and robust.

We next consider the scenario where $X$ is a two-dimensional random Gaussian distribution in Models I.6-I.7. A similar data generation mechanism was followed in~\cite{zhan:22}, who discuss the nonlinear sufficient dimension reduction for distributional objects. For the remaining scenarios, there are no competitive approaches to compare our method with since the proposed global nonlinear Fr\'echet regression method (GNLFR) can accommodate a variety of predictors residing in general metric spaces.\\

\noindent \textbf{Model I.6 (Multivariate distributions as predictors)}:  $\mu\lo Y|X \sim N( \exp(W_2(X,\mu \lo 1)),\nu \lo 1 \hi 2)$ and $\sigma\lo Y|X = 0.1,$ with $\mu \lo 1 \sim N((-1,0)\hi \top, \diag(1,0.5))$.\\

\noindent \textbf{Model I.7 (Multivariate distributions as predictors)}:  $\mu\lo Y|X \sim N( \exp(W_2(X,\mu \lo 1)),\nu \lo 1 \hi 2)$ and $\sigma\lo Y|X = \tau_1\hi \top\Lambda\tau_2,$ with $\mu \lo 1 \sim N((-1,0)\hi \top, \diag(1,0.5))$; $\tau_1 = (1/\sqrt{2}, 1/\sqrt{2})\hi \top,$ $\tau_2 = (1/\sqrt{2}, -1/\sqrt{2})\hi \top,$ $\Lambda =\diag(\lambda\lo 1,\lambda\lo 2)$, where $(\lambda\lo 1, \lambda\lo 2)|X \sim$\\ $N(W_2(X,\mu \lo 2)(1,1)\hi \top, 0.25I_2)$, $\mu \lo 2 \sim N((0,1)\hi \top, \diag(0.5,1))$.\\

When computing $W_2(X,\mu \lo 1)$ and $W_2(X,\mu \lo 2)$, we use the following explicit representations of the Wasserstein distance between two Gaussian distributions:
\begin{align}
	\label{eq:wass:multDim}
	W_2^2(N(m\lo 1,\Sigma\lo 1), N(m\lo 2,\Sigma\lo 2)) = \| m\lo 1 -m\lo 2\|^2 + \| \Sigma\lo 1\hi{1/2} -\Sigma\lo 2 \hi{1/2} \|\lo F^2,
\end{align}
Table~\ref{tab:sim:sceI3} shows a lower MPE for the less complex setting I.6, while the performance of the method improves for higher $n,m$ as before.
\begin{table}[!htb]
	\centering
	\caption{Performances of the proposed global nonlinear Fr\'echet regression for univariate distributional responses with multivariate distributions as predictors under Models I.6-I.7. The lowest number in a row corresponding to each data-generating mechanism is highlighted.}
	\label{tab:sim:sceI3}
	\centering
	\begin{tabular}{|c||cc||cc|}
		\hline
		& \multicolumn{2}{c||}{I.6} & \multicolumn{2}{c|}{I.7} \\ \hline
		$n$\textbackslash{}$m$ & \multicolumn{1}{c|}{50} & 100 & \multicolumn{1}{c|}{50} & 100 \\ \hline
		200 & \multicolumn{1}{c|}{\begin{tabular}[c]{@{}c@{}}0.619 (0.110)\end{tabular}} & \begin{tabular}[c]{@{}c@{}} \textbf{0.534} (0.100)\end{tabular} & \multicolumn{1}{c|}{\begin{tabular}[c]{@{}c@{}}0.719 (0.142)\end{tabular}} & \begin{tabular}[c]{@{}c@{}} \textbf{0.578} (0.131)\end{tabular} \\ \hline
		400 & \multicolumn{1}{c|}{\begin{tabular}[c]{@{}c@{}}0.467 (0.091)\end{tabular}} & \begin{tabular}[c]{@{}c@{}}\textbf{0.388} (0.092)\end{tabular} & \multicolumn{1}{c|}{\begin{tabular}[c]{@{}c@{}}0.635 (0.110)\end{tabular}} & \begin{tabular}[c]{@{}c@{}} \textbf{0.541} (0.112)\end{tabular} \\ \hline
	\end{tabular}
\end{table}

In Model I.8, Hilbertian random functions are taken as predictor objects coupled with univariate distribution responses, where the distribution of the response varies conditional on the predictor values as before.\\

\noindent \textbf{Model I.8 (Random functions as predictors)}:
The predictor trajectories $X$ and associated noisy measurements were generated as follows. Suppose that the simulated process $X$ has the mean function $\mu\lo X(s) = s + \sin(s)$, with covariance function constructed from two eigenfunctions, $\phi\lo 1(s) = \sqrt2\sin(2\pi ks)$ and $\phi\lo 2(s) =  \sqrt2\cos(2\pi ks),$ $0 \leq s \leq 1.$ We chose $\lambda\lo 1 =1,\lambda\lo 2 =0.7$ and $\lambda_k =0$ for $k\geq 3,$ as eigenvalues, and the FPC scores $\xi\lo k$; $(k=1,2)$ were generated from $N(0,\lambda_k)$. Using the Kerhunen-Lo\'eve expansion the predictor process is then given by $X(s) = \mu\lo X(s) + \sum_{k=1}^{\infty} \xi\lo k \phi\lo k(s)$. To adequately reflect both a dense design and an irregular/sparse measurement paradigm, we assume that there is a random number $N_i$ of random measurement times for $X_i$ for the $i\hi{\text{th}}$ subject, which are denoted as $S_{i1},\dots, S_{iN_i}$ and contaminated with measurement errors $\epsilon\lo {ij}$, $1 \leq j \leq N_i$, $1 \leq i \leq n.$ The errors are assumed to be \emph{i.i.d.} with $E(\epsilon\lo {ij}) = 0$ $E[\epsilon\lo {ij}^2] = \sigma\lo X\hi 2 = 0.1$, and independent of functional principal component scores $\xi_{ik}$ that satisfy $E[\xi_{ik}]= 0$, $E[\xi_{ik}\xi_{ik'}] =0$ for $k\neq k'$, and $E[\xi_{ik}^2] = \lambda_k.$ Thus, for the $i\hi{\text{th}}$ sample, the predictor measurement with noise is represented as
$U\lo {ij} =  \mu\lo X(S\lo {ij}) + \sum_{k=1}^{\infty} \xi_{ik} \phi\lo k(S\lo {ij}) + \epsilon\lo {ij},$ $i=1,\dots,n,\ j=1,\dots,N_i$. The data generation mechanism above is similar to~\cite{yao:05} and both a sparse and a dense grid of observation are considered with $N_i = 50$ and $N_i\in\{3,\dots,5\}$, respectively.
Finally, the response as a univariate distribution is constructed as $Y\sim N(\mu\lo Y,\sigma\lo Y)$, and the auxiliary parameters conditional on $X(\cdot)$ are generated independently as $\mu\lo Y|X\sim N((\xi_1,\xi_2)\hi \top \diag(\lambda\lo 1,\lambda\lo 2) (1,-1), \nu \lo 1 \hi 2)$ and $\sigma\lo Y|X = 0.1$.\\

Again, it is evident from Table~\ref{tab:sim:sceI4}, that the method yields better prediction error when the sample size and number of discrete observations per sample in the response is high, favorable for the dense design paradigm for the predictor functions.

\begin{table}[!htb]
	\centering
	\caption{Performances of the proposed global nonlinear Fr\'echet regression (GNLFR) for univariate distributional responses with Hilbertian objects as predictors under Model I.8. The lowest number in a row corresponding to each data-generating mechanism is highlighted.}
	\label{tab:sim:sceI4}
	\centering
	\begin{tabular}{|c||cc||cc|}
		\hline
		& \multicolumn{2}{c||}{\begin{tabular}[c]{@{}c@{}}I.8 (dense design)\end{tabular}} & \multicolumn{2}{c|}{\begin{tabular}[c]{@{}c@{}}I.8 (sparse design)\end{tabular}} \\ \hline
		$n$\textbackslash{}$m$ & \multicolumn{1}{c|}{50} & 100 & \multicolumn{1}{c|}{50} & 100 \\ \hline
		200 & \multicolumn{1}{c|}{\begin{tabular}[c]{@{}c@{}}0.334(0.051)\end{tabular}} & \begin{tabular}[c]{@{}c@{}} \textbf{0.270} (0.049)\end{tabular} & \multicolumn{1}{c|}{\begin{tabular}[c]{@{}c@{}}0.483 (0.130)\end{tabular}} & \begin{tabular}[c]{@{}c@{}}\textbf{0.379} (0.124)\end{tabular} \\ \hline
		400 & \multicolumn{1}{c|}{\begin{tabular}[c]{@{}c@{}}0.211 (0.031)\end{tabular}} & \begin{tabular}[c]{@{}c@{}}\textbf{0.176} (0.032)\end{tabular} & \multicolumn{1}{c|}{\begin{tabular}[c]{@{}c@{}}0.410 (0.022)\end{tabular}} & \begin{tabular}[c]{@{}c@{}}\textbf{0.347} (0.022)\end{tabular} \\ \hline
	\end{tabular}
\end{table}

\paragraph*{{\bf Scenario 2: Multivariate distribution-on-object regression}}
\label{sec:simu:mult:dist}
We now consider the scenario where both $X$ and $Y$ are bivariate random Gaussian distributional objects.
The construction of the kernel $\ka \lo X$ is done using the sliced $2$-Wasserstein distance, which is obtained by computing the average Wasserstein distance of the projected univariate distributions along
randomly picked directions. To define formally,
\begin{definition}[Sliced Wasserstein metric]
	let $\mu \lo 1$ and $\mu \lo 2$ be two measures in $\ca P \lo p(M)$, the set of Borel probability measures on $(M,\ca B(M))$ that have finite $p-$th moment and
	is dominated by the Lebesgue measure on $\real \hi d$, with $M\subset \real \hi d$, $d> 1$.
	Let $S\hi {d-1}$ be the unit sphere in $\real \hi d$. For $\theta \in S\hi {d-1}$, let $T\lo \theta :\real \hi d \to \real$ be the linear transformation $x \mapsto \langle\theta, x\rangle.$ Further, let $\mu \lo 1 \circ T\lo \theta \hi {-1}$ and $\mu \lo 2 \circ T\lo \theta \hi {-1}$
	be the push-forward measures by the mapping $T\lo\theta$. The sliced $p-$Wasserstein distance between $\mu \lo 1$ and $\mu \lo 2$ is then defined
	by
	\begin{align}
		\label{eq:sliced:wass}
		SW_p(\mu \lo 1,\mu \lo 2) = \left(\int_{S\hi {d-1}} W\lo p^p (\mu \lo 1 \circ T\lo \theta \hi {-1}, \mu \lo 2 \circ T\lo \theta \hi {-1})d\theta\right)^{\frac{1}{p}}.
	\end{align}
\end{definition}
For $p = 2$,~\cite{kolo:16} show that the square of sliced Wasserstein distance is conditionally negative definite and hence that the Gaussian RBF kernel defined as $\kappa\lo X(x,x') = \exp(-\gamma\lo X SW_2^2(x,x'))$ is a positive definite kernel.

We generate discrete observations for the predictor distributions $X_i;\ i = 1,\dots,n,$ given by $\{X\lo {ij}\}_{j=1}^m \overset{i.i.d.}{\sim} N(a_i(1,1)\hi \top, b_iI_2),$ where $a_i \overset{i.i.d.}{\sim} N (0.5, 0.5\hi 2)$ and $b_i\overset{i.i.d.}{\sim} Beta(2, 3).$
To compute the Gram matrix associated with the multivariate predictor distribution supported on $M \subset\real \hi d$, $d>1$, the sliced Wasserstein distance is estimated using a Monte Carlo method:
\begin{align*}
	SW_2(\mu\lo {X\lo i},\mu\lo {X\lo k}) \approx \left(\frac{1}{L} \sum_{l=1}^L W_2^2(\mu\lo {X\lo i} \circ T\lo \theta \hi {-1}, \mu\lo{X\lo k} \circ T\lo \theta \hi {-1})\right)\hi{1/2},
\end{align*}
where $ \mu\lo{X\lo i} = \frac{1}{m}\sum_{j=1}^m \delta\lo{X\lo {ij}}$ is the empirical measure for the $i-$th sample, $i=1,\dots,n,$ $\{\theta\lo l\}\lo{l=1}\hi L$ are \emph{i.i.d.} samples drawn from the uniform distribution on $S\hi {d-1} \subset \real\hi d$. The approximation error depends on the number of Monte Carlo samples $L$. In our simulation settings, we set $L = 50.$

The random responses $Y = N (\mu\lo Y , \Sigma\lo Y)$, where $\mu\lo Y\in \real\hi 2$ and $\Sigma\hi Y\in \real\hi{2\times 2}$ are then generated according to the following models.\\

\noindent \textbf{Model II.1 (Multivariate distributions as predictors)}: $\mu\lo Y|X \sim N(W_2(X, \mu \lo 1 )(1, 1)\hi \top, I_2 )$ and $\Sigma\lo Y|X = \diag(1,1).$\\

\noindent \textbf{Model II.2 (Multivariate distributions as predictors)}: $\mu\lo Y|X \sim N(W_2(X, \mu \lo 1 )(1, 1)\hi\top , I_2 )$ and $\Sigma\lo Y|X = \Gamma\Lambda\Gamma\hi \top,$ where $\Gamma = \begin{pmatrix}
	1/\sqrt{2} & 1/\sqrt{2}\\
	-1/\sqrt{2} & 1/\sqrt{2}
\end{pmatrix},$ $\Lambda =\diag(\lambda\lo 1,\lambda\lo 2)$ with $(\lambda\lo 1,\lambda\lo 2)|X \overset{i.i.d.}{\sim} tGamma(W_2^2(X,\mu \lo 2),W_2(X,\mu \lo 2), (0.2,2)),$
where $\mu \lo 1$ and $\mu \lo 2$ are two fixed measures defined by
$\mu \lo 1 = N ((-1, 0)\hi \top, \diag(1, 0.5))$ and $\mu \lo 2 = N ((0, 1)\hi \top, \diag(0.5, 1))$,
and\\ $tGamma(\alpha, \beta, (r_1,r_2 ))$ is the truncated gamma distribution on range $(r_1, r_2)$ with shape
parameter $\alpha$ and rate parameter $\beta.$
The Wasserstein distance between the bivariate Gaussian distributions is computed as per~\eqref{eq:wass:multDim}.\\

If the dimension $d$ of the random probability measures is more than $1$, one does not have an analytic form for the barycenter, and the optimization algorithms to obtain it are complex, in contrast to the case $d =1,$ where the quantile representation of Wasserstein distance leads to an explicit solution via the $L^2$ mean of the quantile functions. The computation of Wasserstein barycenters in multidimensional Euclidean space has been intensively studied (e.g.,~\cite{rabi:12, alva:16,dvur:18, peyr:19}, and one of the most popular methods utilize the Sinkhorn divergence~\citep{cutu:13}, which is an entropy-regularized version of the Wasserstein distance that allows for computationally efficient solutions of the barycenter problem, however at the cost of introducing a bias, as is common for regularized estimation. Due to the gain in efficiency, we adopt this approach in our implementations using the R package \emph{WSGeometry}~\citep{package:hein21}.

Using the same choices for $n$, $m$, and the tuning parameters, we again split the data into a training set and a test set. We use the training set to implement the proposed object regression method at the output predictor points to predict the response in the test set. The whole process is repeated $B= 100$ times, and the prediction error computed between the observed and predicted bi-variate distributional responses in the test set using the average Sliced Wasserstein distance between them, as per~\eqref{eq:sliced:wass}. The averages and standard errors are shown in Table~\ref{tab:sim:sceII}, where a similar pattern of decreased MPE for larger sample size and denser observation grid for the paired sample of distribution is noted.
\begin{table}[!htb]
	\centering
	\caption{Performances of the proposed global nonlinear Fr\'echet regression (GNLFR) under Models II.1-II.2 in Scenario 2. The lowest number in a row is highlighted across different model settings.}
	\label{tab:sim:sceII}
	\centering
	\begin{tabular}{|c||cc||cc|}
		\hline
		& \multicolumn{2}{c||}{II.1} & \multicolumn{2}{c|}{II.2} \\ \hline
		$n$\textbackslash{}$m$ & \multicolumn{1}{c|}{50} & 100 & \multicolumn{1}{c|}{50} & 100 \\ \hline
		200 & \multicolumn{1}{c|}{\begin{tabular}[c]{@{}c@{}}0.620  (0.134)\end{tabular}} & \begin{tabular}[c]{@{}c@{}} \textbf{0.442} (0.130)\end{tabular} & \multicolumn{1}{c|}{\begin{tabular}[c]{@{}c@{}}0.811  (0.200)\end{tabular}} & \begin{tabular}[c]{@{}c@{}} \textbf{0.693}  (0.177)\end{tabular} \\ \hline
		400 & \multicolumn{1}{c|}{\begin{tabular}[c]{@{}c@{}}0.319  (0.094)\end{tabular}} & \begin{tabular}[c]{@{}c@{}}\textbf{0.178} (0.092)\end{tabular} & \multicolumn{1}{c|}{\begin{tabular}[c]{@{}c@{}}0.543  (0.160)\end{tabular}} & \begin{tabular}[c]{@{}c@{}}\textbf{0.329} (0.152)\end{tabular} \\ \hline
	\end{tabular}
\end{table}

\paragraph*{{\bf Scenario 3: SPD matrix object-on-object regression}}
\label{sec:simu:spd}
A common type of random object encountered in brain imaging studies is functional connectivity correlation
matrices, which are positive semi-definite symmetric matrices. Let $(\Omega\lo Y, d\lo F)$ be the space of $r \times r$ symmetric positive definite (SPD)
matrices endowed with Frobenius distance $d\lo F(Y_1, Y_2) = \|Y_1  - Y_2\|\lo F$ as defined in~\eqref{frob:dist} in Section~\ref{sec:exist}.
Two simulation scenarios are considered as follows.\\

\noindent \textbf{Model III.1 (Euclidean predictors)}:
The real-valued predictors $X_i$ are independently sampled from a $Beta(1/2,2)$, while the SPD matrix responses $Y_i$ conditional on $X_i$ are generated according to the model $Y_i = \tilde{Y}_i\tilde{Y}_i\hi \top$, with $\tilde{Y}_i|X_i = \mu(X_i) + [\Sigma(X_i)]\hi{-1/2} Z_i,$ where
for a fixed dimension $r$, the mean vector $\mu(x)$ has components $\mu_j(x) = b_j - 2(x- c_j)^2$, $j= 1,\dots,r$. Here $b_j \sim U(2,4)$ and $c_j \sim U(0, 1)$, and $Z_i$ are sampled independently of $X_i$ as a standard $r-$dimensional Gaussian random vector. the covariance $\Sigma(x)$ is formed by generating a $r \times r$ matrix $A$ with independent $N(0, 0.5)$ random variables in each entry,
then computing $S = 0.5(A + A\hi \top)$. A second $r \times r$ matrix $V$ is generated with elements drawn independently as $U(0, 0.5)$, from which $\theta = 0.5(V + V\hi \top)$ is computed. Finally, with $Exp$ denoting matrix exponentiation and $\odot$ the Hadamard
product, we form $\Sigma(x) = (x + 2x^3) Exp[S \odot \sin(2\pi \theta(x + 0.1))].$\\

\noindent \textbf{Model III.2 (SPD matrix objects as predictors)}:
The predictors are now themselves SPD matrices. This is generated as the covariance matrix computed from a  $p$-variate Gaussian random vector with independent components each with mean $0$ and variance $1$ for each sample. The predictors are projected down on a desired direction vector $\beta$ whose each component $\beta_j\sim U(0,1)$, $j=1,\dots,p$ to compute $\tilde{X}_i = X_i\beta$. Here, we choose $p = 5.$ Now the response matrices are generated as before in Model III.2 conditional on $\tilde{X}_i$.\\

In order to apply the proposed method, again the Gaussian RBF kernel given by $\ka\lo X(x,x') = \exp(-\gamma\lo Xd^2\lo F(x,x')$ is taken to compute the Gram matrix in the predictor space, with the tuning parameter chosen as before. From a sample $(X_i,Y_i)_{i=1}^n$ the minimization in~\eqref{estimator_final} can be reformulated by
setting $\hat{f}\lo\oplus(x) = \frac{1}{n} \sum_{i=1}^n w_{in}(x)Y_i$ and computing the correlation matrix which is nearest to the matrix $\hat{f}\lo\oplus(x)$, which is implemented by the alternating projections algorithm via the \emph{nearPD()} function in the \emph{Matrix} R package.

We compare performances of the proposed method for a combination of sample size and the dimension of the response matrices given by $n$ and $r$, respectively, by computing the Frobenius distance between the true and the predicted SPD matrix responses in the test set, using the model fit on the training set, as described before. The first two columns of Table~\ref{tab:sim:sce34} display the average prediction error across $100$ replications of the above process. Our method fares better for increased sample size, while the dimension of the response SPD matrices is lower in both simulation scenarios.

\begin{table}[!htb]
	\centering
	\caption{Performances of the proposed global nonlinear Fr\'echet regression (GNLFR) under Models III.1-III.2 and IV.1 in Scenarios 3 and 4. The lowest number in a row is highlighted across different model settings.}
	\label{tab:sim:sce34}
	\centering
	\begin{tabular}{|c||cc||cc||cc|}
		\hline
		& \multicolumn{2}{c||}{III.1} & \multicolumn{2}{c||}{III.2} & \multicolumn{2}{c|}{IV.1} \\ \hline
		$n$\textbackslash{}$r$ & \multicolumn{1}{c|}{5} & 20 & \multicolumn{1}{c|}{5} & 20 & \multicolumn{1}{c|}{5} & 20 \\ \hline
		200 & \multicolumn{1}{c|}{\begin{tabular}[c]{@{}c@{}}\textbf{0.119} \\ (0.041)\end{tabular}} & \begin{tabular}[c]{@{}c@{}}0.275 \\ (0.040)\end{tabular} & \multicolumn{1}{c|}{\begin{tabular}[c]{@{}c@{}}\textbf{0.226} \\ (0.130)\end{tabular}} & \begin{tabular}[c]{@{}c@{}}0.786 \\ (0.110)\end{tabular} & \multicolumn{1}{c|}{\begin{tabular}[c]{@{}c@{}}\textbf{0.161}\\  (0.011)\end{tabular}} & \begin{tabular}[c]{@{}c@{}} 0.235 \\ (0.031) \end{tabular} \\ \hline
		400 & \multicolumn{1}{c|}{\begin{tabular}[c]{@{}c@{}}\textbf{0.048} \\ (0.037)\end{tabular}} & \begin{tabular}[c]{@{}c@{}}136 \\ (0.035)\end{tabular} & \multicolumn{1}{c|}{\begin{tabular}[c]{@{}c@{}}\textbf{0.127} \\ (0.110)\end{tabular}} & \begin{tabular}[c]{@{}c@{}}0.502 \\ (0.097)\end{tabular} & \multicolumn{1}{c|}{\begin{tabular}[c]{@{}c@{}}\textbf{0.079} \\ (0.012)\end{tabular}} & \begin{tabular}[c]{@{}c@{}}0.145 \\ (0.029)\end{tabular} \\ \hline
	\end{tabular}
\end{table}

\paragraph*{{\bf Scenario 4: Network object-on-object regression}}
\label{sec:simu:netwrok} \ \\ 
\noindent \textbf{Model IV.1 (Euclidean predictors)}:  Let $G = (V, E)$ be a simple (with no self-loops), weighted, undirected network with a set of nodes $V = \{v_1, \dots, v_r\}$ and a set of edge weights $E = \{w\lo {ij}: w\lo {ij}\geq 0,\ i, j = 1,\dots, r\}$, where $w\lo {ij} = 0$ indicates $v_i$ and $v_j$ are not connected and $w\lo {ij}>0$ otherwise, with $w\lo {ij} <M$ for some $M>0.$ A network can be uniquely represented by its graph Laplacian $L = (l\lo {ij}),$ where $	l\lo {ij} = -w\lo {ij} \text{ if } i \neq j$ and $l\lo {ij} = \sum_{k\neq i} w_{ik} \text{ if } i = j,$
for $i, j = 1,\dots, r$.
The space of graph Laplacians is given by  $\mathcal{L}_r = \{L = (l\lo {ij}) : L = L\hi \top, \ L1_r = 0_r,\ -W\leq l\lo {ij} \leq 0\ \text{ for some W }\geq 0 \text{ and } i\neq j\},$ where $1_r$ and $0_r$ are the $r$-vectors of ones and zeroes, respectively.
Note that $\mathcal{L}_r$ is not a linear space, but a bounded, closed, and convex subset in $\real \hi {r^2}$ of dimension $r(r-1)/2$. Owing to the fact that $x\hi \top Lx \geq 0$ for all $x \in \real \hi r$ and $L\in \mathcal{L}_r$, it can be seen as a metric space of positive-semidefinite matrix objects, equipped with a suitable choice of metric such as the Frobenius or power metric.

To assess the performance of our proposed methods, we consider the space $(\mathcal{L}_r, d\lo F)$, where $d\lo F$ is the Frobenius metric as per~\eqref{frob:dist}. The data generation mechanism, as follows, is similar to that in~\cite{zhou:22}. Denote the half vectorization
excluding the diagonal of a symmetric and centered matrix by $vech$, with inverse operation $vech\hi{-1}$. By the symmetry and centrality, every graph Laplacian $L$ is fully known by its upper (or lower) triangular part, which can then be vectorized into $vech(L)$, a vector of length $d = r(r -1)/2$. We construct the conditional distributions $F_{L|X}$ by assigning an independent beta distribution to each element of $vech(L)$. Specifically, a random
sample $(\beta_1, \dots, \beta_d)\hi \top$ is generated using beta distributions whose parameters depend on the scalar predictor $X$ and vary under different simulation scenarios. The random response $L$
is then generated conditional on $X$ through an inverse half vectorization $vech\hi{-1}$ applied to $(\beta_1, \dots, \beta_d)\hi \top$. The the true
regression function $m(x)$ is defined as $m(x) = vech\hi{-1}(-x,\dots, -x)$, $L = vech\hi{-1}(\beta_1, \dots, \beta_d)\hi \top$, where $\beta_j\overset{i.i.d.}{\sim} Beta(X, 1 - X).$
To ensure that the random response $L$ generated in simulations resides in $\mathcal{L}_r$, the off-diagonal entries $-\beta_j$ $j = 1,\dots,d,$ need to be nonpositive and bounded below. Thus we choose $\beta_j\overset{i.i.d.}{\sim} Beta(X, 1-X)$. The scalar predictor $X_i$ are randomly sampled from a $Unif(0,1)$ distribution to obtain the samples of pairs $(X_i, L_i)$, $i=1,\dots,n$, setting $r = 5, 20$, and following the above procedure.
The prediction error w.r.t the Frobenius metric is shown in the rightmost column of Table~\ref{tab:sim:sce34}. The method performs better for higher $n$ and lower $r.$

\section{A real application}
\label{sec:data}
\begin{figure}[!htb]
	\centering
	\begin{subfigure}{0.45\textwidth}
		\includegraphics[width=.9\textwidth]{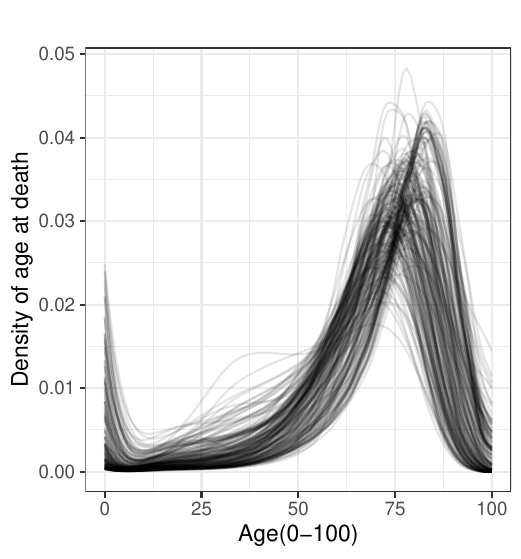}
		%	\caption{image1}
		%	\label{fig:1}
	\end{subfigure}\hfil % <-- added
	\begin{subfigure}{0.45\textwidth}
		\includegraphics[width=.9\textwidth]{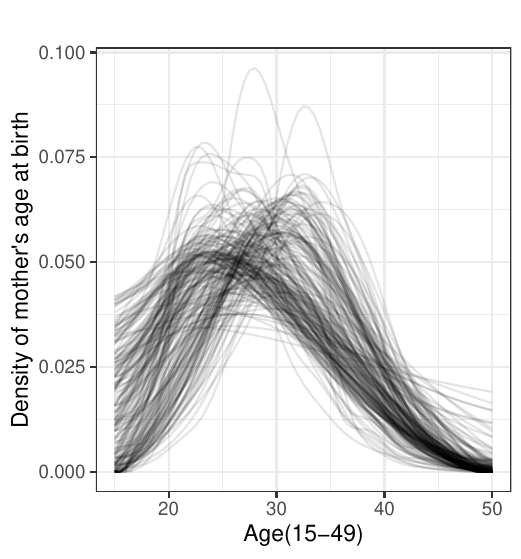}
		%\caption{image2}
		%\label{fig:2}
	\end{subfigure}
	\centering
	\caption{Visualization of distributional objects represented as densities of age at death and mother's age at birth for a sample of 194 countries.}
	\label{fig:data:mort:densities}
\end{figure}

In this application, we explore the relationship between the distribution of age at death and
that of the mother’s age at birth at a country level. Going beyond summary statistics such as mortality or fertility rate, viewing the entire distributions as samples of data is more informative and insightful for understanding the nature of human longevity and its dependence on relevant predictors. The data was obtained from the UN World Population Prospects $2019$ Databases (\url{https://population.un.org}). For this analysis, we focused on $n = 194$ countries over the period of time $2015-2020$. The mortality data was available in the form of life tables over the age interval  $[0,110]$ (all in years), while the number of births was categorized by the mother’s age every five years over the age bracket $[15,50]$. We used bin widths equal to 5 years to construct the histograms for the mortality and fertility distributions, respectively,  and proceeded to obtain the smooth densities by applying local linear regression using the \emph{frechet} package~\citep{fr:package} at the country level, with the domains of the age-at-death and mother's age-at-birth densities as $[0,110]$ and $[15,50]$ years, respectively. The densities were assumed to lie in the space of univariate distributions equipped with the Wasserstein metric $(\Omega\lo Y, d\lo W)$ in~\eqref{wass:dist}. Figure~\ref{fig:data:mort:densities} shows the sample of densities as observed.

\begin{figure}[!htb]	
	\centering
	\includegraphics[width=\textwidth]{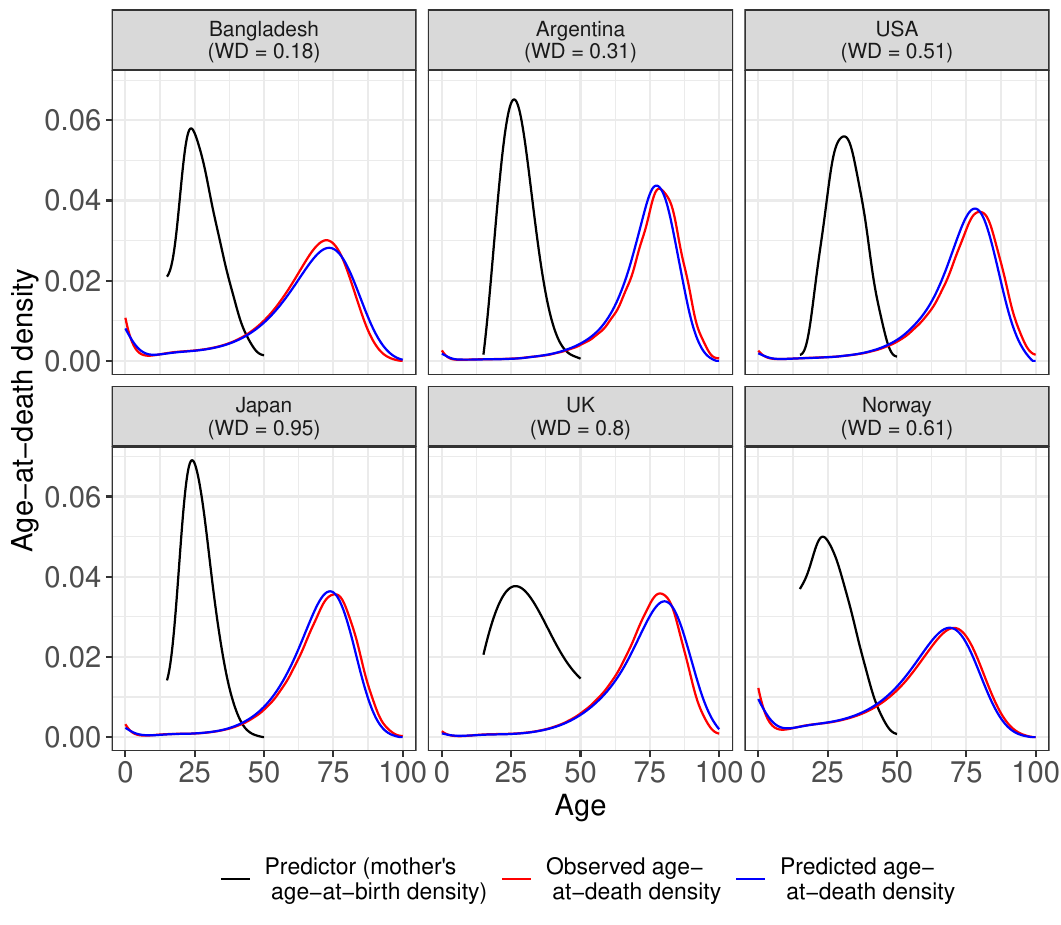}
	\centering
	\caption{Visualization of distributional objects represented as densities of age at death and mother's age at birth for a sample of 194 countries.}
	\label{fig:data:mort:obs_pred}
\end{figure}

We applied the proposed nonlinear object-on-object regression method with age-at-death densities as responses and mother's age-at-birth densities as predictors to compare the evolution of mortality distributions among different countries aggregated for the calendar years $2015-2020.$ We show the densities obtained from a leave-one-out prediction results (in blue) together with the observed distributional responses (in red) Figure~\ref{fig:data:mort:obs_pred} for a select few countries, which showcases different patterns of mortality change over changes in the predictor distribution. The predictor densities of the mother's age at birth are also overlaid in the same panel of plots. The Wasserstein distance discrepancy (WD) between the observed and predicted distributions is also shown.
Specifically, we selected the countries Bangladesh, Argentina, the USA, Japan, the UK, and Norway, ordered by the lowest to the highest value of the mode of the mother's age-at-death densities. Both the observed and predicted age-at-death densities across the panels from left to right are seen to be more right-shifted, indicating increased longevity corresponding to a higher age at birth for the mother.
Further, for Japan, Norway, and the USA,
the rightward mortality shift is seen to be more pronounced than suggested by the prediction, indicating that longevity extension is more than anticipated, while the mortality distribution for the UK seems to shift to the right at a slower pace than predicted, leading to a relatively larger WD with a value of $0.8$ between the observed and predicted
response. In contrast, the regression fit for Argentina and Bangladesh is quite accurate.

\begin{figure}[!htb]
	\begin{subfigure}{0.448\textwidth}
		\centering
		\includegraphics[width=.9\textwidth]{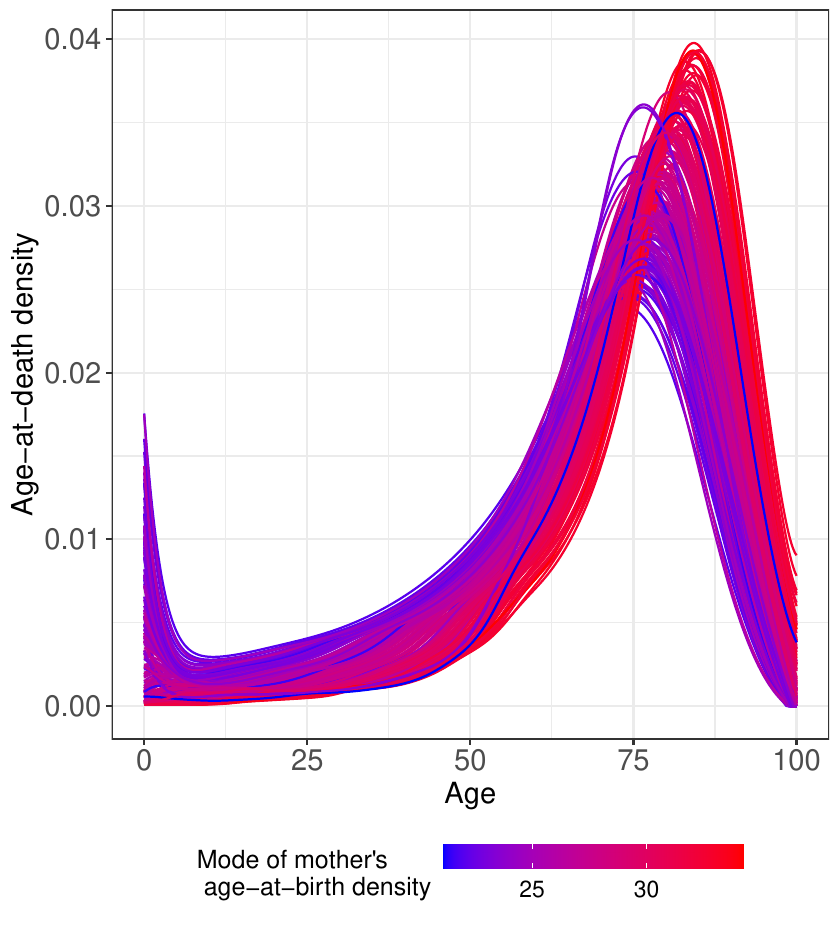}
		\centering
		\caption{The changes in the density of the age-at-death distribution as the mode of the distribution of the mother's age-at-birth ranges from low (blue) to high (red) are displayed.}
		\label{fig:data:mort:changeZ}
	\end{subfigure}\hfil % <-- added
	\begin{subfigure}{0.51\textwidth}
		\centering
		\includegraphics[width=\textwidth]{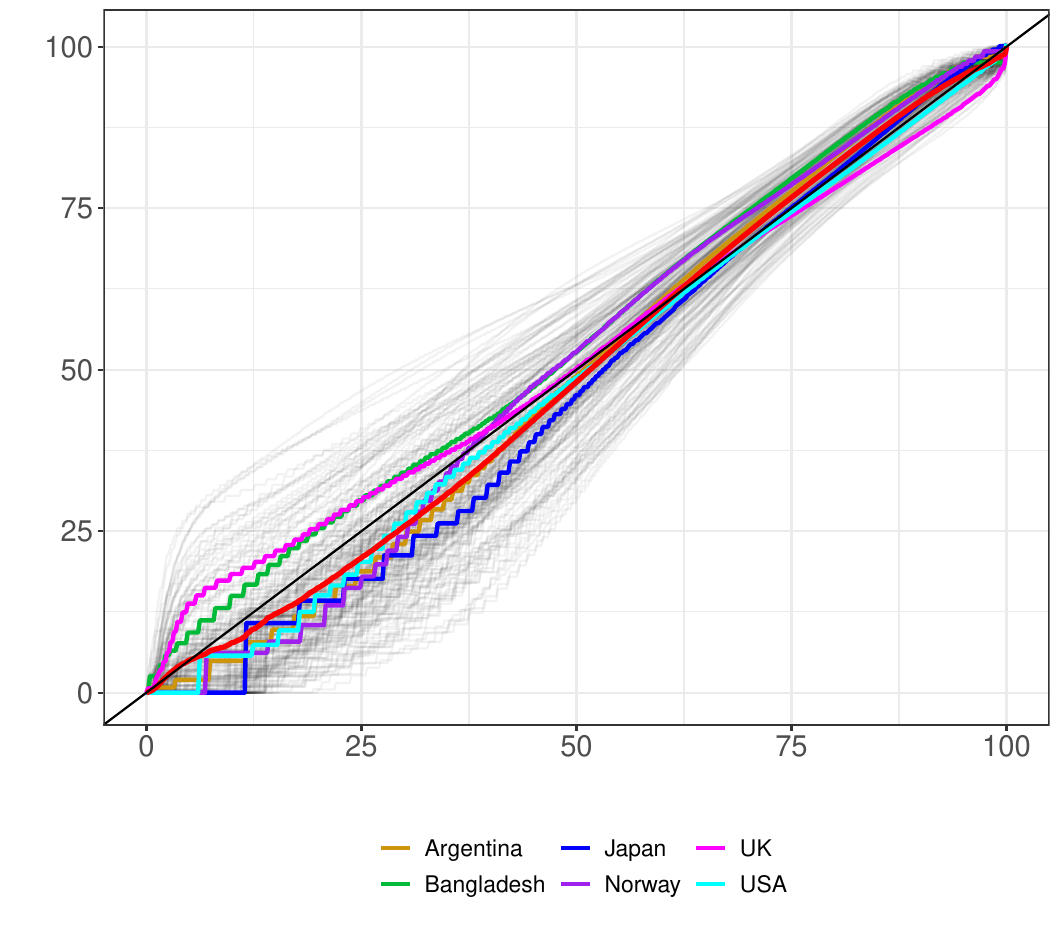}
		\centering
		\caption{Residual maps corresponding to $n= 194$ countries are plotted in gray, with specific countries highlighted. The identity map and the average residual map are overlaid in black and red, respectively.}
		\label{fig:data:mort:resi}
	\end{subfigure}
	\caption{Visualization of the effect of the mother's age-at-birth and residual maps. }
\end{figure}

The effect of the mother's age-at-birth is elicited in Figure~\ref{fig:data:mort:changeZ}, where the model is fitted for varying levels of the mode of the predictor distribution. The fitted densities are color-coded such that blue to red indicates smaller to larger values of the mode of the age-at-birth densities.
We find that lower age-at-birth of the mother is associated with left-shifted age-at-death distributions in general, while modes at higher age-at-birth correspond to a shift of the mode of the age-at-death toward the right. Child mortality is associated with low and high values of age-at-birth for the mother, which concurs with the observations made earlier.

The fit of the model is further demonstrated by computing the estimation error by virtue of the residual map for the $i\hi{\text{th}}$ subject given by $T_i:\Omega\lo Y\to \Omega\lo Y$, which is defined as the optimal transport map $T_i =  \nu_i\#\hat{\nu}_i,$ that pushes forward the observed response $\nu_i$ to the fitted value $\hat{\nu}_i$. Using the theory of optimal transport for univariate distributions~\citep{vill:09}, this map can be explicitly computed as $T_i = Q_{\hat{\nu}_i}\circ F_{\nu_i}$, where  $Q_{\hat{\nu}_i}$ and $F_{\nu_i}$ are, respectively, the quantile function and the CDF of the distributions $\hat{\nu}_i$ and $\nu_i$.
Using these residual maps, one can obtain an analog of the ``residual plot'' in the classical regression case, compared to the identity map. Looking at the deviation from the identity map, one can see in which parts of the support of the distributions the model provides a good fit and where less so, and the departure from the identity can serve as a diagnostic tool for the validity of the model. Note
that, contrary to classical regression, where the residuals add up to zero by construction, the residual
maps are not constrained to have a mean equal to the identity.

The residual maps computed for each of the $194$ countries are plotted in Figure~\ref{fig:data:mort:resi}. One can see that the pointwise variability is much more prominent for younger ages and decreases for progressively older ages, indicating many other plausible factors affecting mortality at younger ages. The identity map is overlaid in black. The mean transport map for the residuals, plotted in red, lies very close to the identity map, which provides evidence in support of the validity of our model. The residual maps of the specific countries considered in Figure~\ref{fig:data:mort:obs_pred} are highlighted. Similar patterns of right-shifted distributions, especially near the age-at-death $[15,40]$ years, are observed for the highlighted countries. For example, while the
evolution of the mortality distributions for Japan and the USA can be viewed as mainly a rightward
shift over calendar years; this is not the case for the UK, where compared with the fitted response, the
actual rightward shift of the mortality distribution seems to be accelerated for those above age 65 and decelerated for those below age 65.

To evaluate the out-of-sample prediction performance of the method, we randomly split the dataset
into a training set and a test set, and use the fits obtained from the training set to predict the responses to the test set using only the predictors present in the test set. As a measure of the efficacy of the fitted model, we compute the mean prediction error as the Wasserstein discrepancy between the observed and the predicted distributions in the test set. We repeat the process $100$ times to obtain the average prediction error, which comes out low ($0.693$ with a standard error of $0.151)$, supporting the efficacy of the model.

\section{Discussion}
\label{sec:concl}
In this contribution, we proposed a nonlinear global object-on-object regression method based on the intrinsic geometry of the metric space where the responses reside coupled with suitable linear operators defined via the reproducing kernel Hilbert space on the predictor space. This contribution is one of the first to model the regression relationship between metric-valued object pairs beyond scalar-or-vector-valued predictors. Further, we bridge the gap between the conditional Fr\'echet mean, and the globally linear Fr\'echet means proposed by~\cite{pete:19} by introducing the notion of a more general weak conditional Fr\'echet mean. This provides a way to link random object data analysis to non-linear global reproducing kernel Hilbert spaces (RKHS) regression models, allowing for arbitrary non-linear functions beyond linear or polynomial regression.
In the process of defining the weak conditional Fr\'echet mean, the weak conditional moments for the classical Hilbertian objects are discussed, and the relevant properties are proved, which is an important construct on its own and makes a separate contribution to the literature.

The concept of weak Fr\'echet moments can be extended to Fr\'echet median or as a minimizer of Huber loss by substituting $E[d\hi 2\lo Y(Y,\cdot)\bbar X]$ by $E[\rho\lo Y(Y,\cdot)\bbar X]$, for any appropriate convex loss function $\rho\lo Y$ in the metric space $(\Omega\lo Y,d\lo Y)$, depending on the context and interpretation of the problem. This calls for potential future research. The selection of a suitable metric in the response or predictor space is also an open problem.

Further, the rate of convergence of the proposed estimator is derived as $\approx n^{-1/4},$ which entails from the work of~\cite{li:17}. This rate can be further improved using a suitable rate carried out from the RKHS regression literature.

\appendix
\section{Technical assumptions for M-estimators}
\begin{assumption}
	\label{ass:exists} The weak conditional Fr\'echet means $f\lo\oplus(x)$ and $\hat{f}\lo\oplus(x)$ exist and are unique, the latter almost surely. Further, the minimizer at the population level is well separated. i.e.,
	for any $\epsilon>0,$
	\[
	\underset{d\lo Y(y,f\lo\oplus(x))>\epsilon}{\inf \ } J(y,x) - J(f\lo\oplus(x),x) >0.
	\]
\end{assumption}
\begin{assumption}
	\label{ass:entropy}
	Let $B_\delta(f\lo\oplus(x)) \subset \Omega\lo Y$ be the ball of radius $\delta$, centered at $f\lo\oplus(x)$ and $N(\epsilon,B_\delta(f\lo\oplus(x)),d\lo Y)$ be its covering number using balls of radius $\epsilon$. Then the entropy integral is computed from the covering number given by
	\[
	J = J(\delta):= \int_{0}^{1} \sqrt{1 + \log N(\delta\epsilon,B_\delta(f\lo\oplus(x)),d\lo Y)} d\epsilon = O(1) \text{ as } \delta \to 0.
	\]
\end{assumption}
\begin{assumption}
	\label{ass:curv}
	There exist constants $\eta>0$, $C>0$, and $\beta>1$, 
	possibly depending on $x \in (\Omega\lo X,d\lo X )$, such that
	\[
	J(y,x) - J(f\lo\oplus(x),x) \geq C d\lo Y\hi \beta(y,f\lo\oplus(x)),
	\]
	for any small neighborhood $d \lo Y(y,f\lo\oplus(x)) <\eta.$
\end{assumption}

Assumption~\ref{ass:exists} is commonly used to establish the consistency of an M-estimator; see Chapter 3.2 in~\cite{vand:00}. In particular, it ensures that weak convergence of the empirical process $\tilde{J}_{n}$ to the population process $J$, which in turn implies convergence of their minimizers. The conditions on the covering number in Assumption~\ref{ass:entropy} and curvature in Assumption~\ref{ass:curv} arise from empirical process theory and control the behavior of $\tilde{J}_{n} - J$ near the minimum, which is necessary to obtain rates of convergence.
These assumptions are again satisfied for many random objects of interest, the common examples of random objects such as distributions, covariance matrices, networks, and so on (see Propositions 1-3 of \cite{pete:19}).

\bibliographystyle{apalike}
\bibliography{rkhs_reg.bib}	
\end{document}